%% file: DMCM.tex
\documentclass[review]{elsarticle}

\usepackage{lineno,hyperref}
\modulolinenumbers[5]

\usepackage{times}% For figures
\usepackage{graphicx} % more modern
\usepackage{enumerate}
\usepackage{amssymb}
\usepackage{amsmath}
\usepackage{algorithm}
\usepackage{algorithmic}
\usepackage{subcaption}
\usepackage{bm}
\usepackage[framed,numbered]{mcode}
\usepackage{epsfig}
\usepackage{epstopdf}
\usepackage{color}
\newtheorem{theorem}{Theorem}

\newtheorem{definition}{Definition}
\newtheorem{proposition}{Proposition}

\DeclareMathOperator*{\st}{s.t.}
\DeclareMathOperator*{\argmin}{argmin}
\renewcommand{\u}{\mathbf{u}}
\newcommand{\D}{\mathbf{D}}
\newcommand{\V}{\mathbf{V}}
\newcommand{\Vk}{\mathbf{V}_k}

\newcommand{\M}{\mathbf{M}}
\newcommand{\Mk}{\mathbf{M}_k}
\newcommand{\Mkk}{\mathbf{M}_{k+1}}
\newcommand{\norm}[1]{\left\lVert#1\right\rVert}
\input{MACPdef-ams.tex}
\input{MACPdef.tex}

\newcommand{\canyi}[1]{\textcolor{black}{#1}\ }
\newenvironment{CanyiPar}[1]{%
	\leavevmode\color{#1}\ignorespaces%
}{}

\begin{document}
	
	\begin{frontmatter}
		
\title{Optimized Projections for Compressed Sensing via Direct Mutual Coherence Minimization}

%		\tnotetext[mytitlenote]{Fully documented templates are available in the elsarticle package on \href{http://www.ctan.org/tex-archive/macros/latex/contrib/elsarticle}{CTAN}.}
		
		\author{Canyi Lu$^a$, Huan Li$^b$, Zhouchen Lin$^{b,c}$}
				
		\address{$^a$Department of Electrical and Computer Engineering, National University of Singapore, Singapore}
		\address{$^b$Key Laboratory of Machine Perception (MOE), School of EECS, Peking University, China}
		\address{$^b$Cooperative Medianet Innovation Center, Shanghai Jiao Tong University, China}

%		\author[mysecondaryaddress]{Global Customer Service\corref{mycorrespondingauthor}}
%		\cortext[mycorrespondingauthor]{Corresponding author}
%		\ead{support@elsevier.com}
%		
%		\address[mymainaddress]{1600 John F Kennedy Boulevard, Philadelphia}
%		\address[mysecondaryaddress]{360 Park Avenue South, New York}

		\begin{abstract}
			Compressed Sensing (CS) is a new data acquisition theory
			based on the existence of a sparse representation of a signal and
			a projected dictionary $\PP\D$, where $\PP\in\mathbb{R}^{m\times
				d}$ is the projection matrix and $\D\in\mathbb{R}^{d\times n}$ is
			the dictionary. To   recover the signal from a small number
			$m$ of measurements, it is expected that the projected dictionary
			$\PP\D$ is of low mutual coherence. Several previous methods
			attempt to find the projection $\PP$ such that the mutual
			coherence of $\PP\D$ is  low. However, they do not
			minimize the mutual coherence directly and thus they may
			be far from optimal. Their used solvers lack convergence
			guarantee and thus the quality of their   solutions is not
			guaranteed. This work aims to address these issues. We propose to
			find an optimal projection matrix by minimizing the mutual
			coherence of $\PP\D$ directly. This leads to a nonconvex nonsmooth
			minimization problem. We   approximate it by smoothing,
			solve it by alternating minimization and prove the
			convergence of our algorithm. To the best of our knowledge, this
			is the first work which directly minimizes the mutual coherence of
			the projected dictionary and has convergence guarantee. Numerical
			experiments demonstrate that our method can recover
			sparse signals better than existing ones.
		\end{abstract}
		
		
	\end{frontmatter}
	
	\linenumbers

\section{Introduction}

Compressed Sensing (CS)
\cite{candes2006robust,donoho2006compressed} is a new
sampling/data acquisition theory asserting that one can exploit
sparsity or compressibility when acquiring signals of interest. It
shows that signals which have a sparse representation with respect
to appropriate bases can be recovered from a small number of
measurements. A fundamental problem in CS is how to construct a
measurement matrix such that the number of measurements is near
minimal.

Consider a signal $\x\in\mathbb{R}^d$ which is assumed to have a
sparse representation with respect to a fixed overcomplete
dictionary $\D\in\mathbb{R}^{d\times n}$ $(d< n)$. This can be
described as
\begin{equation}\label{eq_dr}
\x=\D\balpha,
\end{equation}
where $\balpha\in\mathbb{R}^n$ is a sparse representation
coefficient, i.e., $\norm{\balpha}_0\ll n$. Here
$\norm{\balpha}_0$ denotes the $\ell_0$-norm which counts the
number of nonzero elements in $\balpha$. The solution to problem
(\ref{eq_dr}) is not unique since $d<n$. To find an appropriate
solution in the solution set of (\ref{eq_dr}), we need to use some
additional structures of $\D$ and $\balpha$. Considering that
$\balpha$ is sparse, we are interested in finding the sparsest
representation coefficient $\balpha$. This leads to the following
sparse representation problem
\begin{equation}\label{eq_sr}
\min_{\balpha}\norm{\balpha}_0,\   \st \  \x=\D\balpha.
\end{equation}
However, the above problem is NP-hard \cite{natarajan1995sparse}
and thus is challenging to solve. Some algorithms, such as Basis
Pursuit (BP) \cite{chen1998atomic} and Orthogonal Matching Pursuit
(OMP) \cite{pati1993orthogonal}, can be used to find suboptimal
solutions.

An interesting theoretical problem is that under what conditions
the optimal solution to (\ref{eq_sr}) can be computed. If the
solution is computable, can it be exactly or approximately
computed by BP or OMP? Some previous works answer the above
questions based on the mutual coherence of the dictionary $\D$
\cite{gribonval2003sparse}.
\begin{definition}
Given $\D=[{\dd}_1,\cdots,\dd_n]\in\mathbb{R}^{d\times n}$, its mutual coherence is defined as the largest absolute and normalized inner product between different columns of $\D$, i.e.,
\begin{equation*}
\mu(\D)=\max_{\substack{1\leq i,j\leq n\\i\neq j}}\frac{|\dd_i^T\dd_j|}{\norm{\dd_i}\norm{\dd_j} }.
\end{equation*}
\end{definition}
The mutual coherence measures the highest correlation between any
two columns of $\D$. It is expected to be as low as possible in
order to find the sparest solution to (\ref{eq_sr}).
\begin{theorem}\label{thm1} \cite{gribonval2003sparse,donoho2003optimally,tropp2004greed}
For problem (\ref{eq_sr}), if $\balpha$ satisfies
\begin{equation}
\norm{\balpha}_0<\frac{1}{2}\left(1+\frac{1}{\mu(\D)}\right),
\end{equation}
then the following results hold:
\begin{itemize}
    \item $\balpha$ is the solution to (\ref{eq_sr}).
    \item $\balpha$ is also the solution to the following convex $\ell_1$-minimization problem
    \begin{equation*}\label{eq_srl1}
    \min_{\balpha}\norm{\balpha}_1,\   \st \  \x=\D\balpha,
    \end{equation*}
    where $\norm{\balpha}_1=\sum_i|\alpha_i|$ is the $\ell_1$-norm of $\balpha$.
    \item $\balpha$ can be obtained by OMP.

\end{itemize}
\end{theorem}
The above theorem shows that if the mutual coherence of $\D$ is
low enough, then the sparest solution to (\ref{eq_sr}) is
computable. Thus, how to construct a dictionary $\D$ with low
mutual coherence is crucial in sparse coding. In CS, to reduce the
number of measurements, we face a similar problem on the sensing
matrix construction.

The theory of CS guarantees that a signal having a sparse
representation can be recovered exactly from a small set of linear
and nonadaptive measurements. This result suggests that it may be
possible to sense sparse signals by taking far fewer measurements
than what the conventional Nyquist-Shannon sampling theorem
requires. But note that CS differs from classical sampling in
several aspects. First, the sampling theory typically considers
infinite-length and continuous-time signals. In contrast, CS is a
mathematical theory that focuses on measuring finite-dimensional
vectors in $\mathbb{R}^n$. Second, rather than sampling the signal
at specific points in time, CS systems typically acquire
measurements in the form of inner products between the signal and
general test functions. At last, the ways to dealing with the
signal recovery are different. Given the signal
$\x\in\mathbb{R}^{d}$ in (\ref{eq_dr}), CS suggests replacing
these $n$ direct samples with $m$ indirect ones by measuring
linear projections of $\x$ defined by a proper projection or
sensing matrix $\PP\in\mathbb{R}^{m\times d}$, i.e.,
\begin{equation}
\y=\PP\x,
\end{equation}
such that $m\ll d$. It means that instead of sensing all $n$
elements of the original signal $\x$, we can sense $\x$ indirectly
by its compressed form $\y$ in a much smaller size $m$.
Surprisingly, the original signal $\x$ can be recovered from the
observed $\y$ by using the sparse representation in (\ref{eq_dr}),
i.e, $\y=\PP\D\balpha$ with the sparest $\balpha$. Thus the
reconstruction requires solving the following problem
\begin{equation}\label{eq_ed}
\min_{\balpha} \norm{\balpha}_0, \ \st \ \y=\M\balpha,
\end{equation}
where $\M=\PP\D\in\mathbb{R}^{m\times n}$ is called the effective
dictionary. Problem (\ref{eq_ed}) is also NP-hard. As suggested by
Theorem \ref{thm1}, if the mutual coherence of $\PP\D$ is low
enough, then the solution $\balpha$ to (\ref{eq_ed}) is computable
by OMP or by solving the following convex problem
\begin{equation}\label{eq_edl1}
\min_{\balpha} \norm{\balpha}_1, \ \st \ \y=\M\balpha.
\end{equation}
Finally, the original signal $\x$ can be reconstructed by
$\x=\D\balpha$. So it is expected to find a proper projection
matrix $\PP$ such that $\mu(\PP\D)$ is low. Furthermore, many
previous works \cite{elad2007optimized,xu2010optimized} show that
the required number of measurements for recovering the signal $\x$
by CS can be reduced if $\mu(\PP\D)$ is low.

In summary, the above discussions imply that by choosing an
appropriate projection matrix $\PP$ such that $\mu(\PP\D)$ is low
enough, the true signal $\x$ can be recovered with high
probability by efficient algorithms. At the beginning, random
projection matrices were shown to be good choices since their
columns are incoherent with any fixed basis $\D$ with high
probability \cite{candes2006near}. However, many previous works
\cite{elad2007optimized,duarte2009learning,xu2010optimized} show
that well designed deterministic projection matrices can often
lead to better performance of signal reconstruction than random
projections do. In this work, we focus on the construction of
deterministic projection matrices. We first give a brief review on
some previous deterministic methods.

\subsection{Related Work}

In this work, we only consider the case that $\D$ is fixed while
$\PP$ can be changed. Our target is to find $\PP$ by minimizing
$\mu(\M)$, where $\M=\PP\D$. If each column of $\M$ is normalized
to have unit Euclidean length, then
$\mu{(\M)}=\norm{\G}_{\infty,\text{off}}$, where
$\G=(g_{ij})=\M^T\M$ is named as the Gram matrix and
$\norm{\G}_{\infty,\text{off}}=\max_{i\neq j}|g_{ij}|$ is the
largest off-diagonal element of $|\G|$. Several previous works
used the Gram matrix to find the projection matrix $\PP$
\cite{elad2007optimized,duarte2009learning,xu2010optimized}. We
give a review on these methods in the following.

\subsubsection{The Algorithm of Elad}

The algorithm of Elad \cite{elad2007optimized} considers
minimizing the $t$-averaged mutual coherence defined as the
average of the absolute and normalized inner products between
different columns of $\M$ which are above $t$, i.e.,
\begin{equation*}
\mu_t{(\M)}=\frac{\sum_{1\leq i,j\leq k, \ i\neq j} \chi_t(|g_{ij}|) |g_{ij}| }{\sum_{1\leq i,j\leq k, \ i\neq j} \chi_t(|g_{ij}|) },
\end{equation*}
where $\chi_t(x)$ is the characteristic function defined as
\begin{equation*}
\chi_t(x) = \begin{cases}
1, \quad \text{if } x\geq t, \\
0, \quad \text{otherwise},
\end{cases}
\end{equation*}
and $t$ is a fixed threshold which controls the top fraction of
the matrix elements of $|\G|$ that are to be considered.

To find $\PP$ by minimizing $\mu_t(\M)$, some properties of the
Gram matrix $\G=\M^T\M$ are used. Assume that each column of $\M$
is normalized to have unit Euclidean length. Then
\begin{equation}
\diag{\G}=\1,
\end{equation}
\begin{equation}
\rank{\G}=m.
\end{equation}
The work \cite{elad2007optimized} proposed to minimize $\mu_t(\M)$
by iteratively updating $\PP$ as follows. First, initialize $\PP$
as a random matrix and normalize each column of $\PP\D$ to have
unit Euclidean length. Second, shrink the elements of $\G=\M^T\M$
(where $\M=\PP\D$) by
\begin{equation*}
g_{ij}=\begin{cases}
\gamma g_{ij}, \quad\quad\quad   \text{\ \ if } |g_{ij}|\geq t,\\
\gamma t\text{sign}{(g_{ij})}, \quad \text{if }t>|g_{ij}|\geq \gamma t,\\
g_{ij}, \quad\quad\quad\quad   \text{ if } \gamma t>|g_{ij}|,
\end{cases}
\end{equation*}
where $0<\gamma<1$ is a down-scaling factor. Third, apply SVD and
reduce the rank of $\G$ to be equal to $m$. At last, build the
square root $\SS$ of $\G$: $\SS^T\SS=\G$, where
$\SS\in\mathbb{R}^{m\times n}$, and find $\PP=\SS\D^\dagger$,
where $^\dagger$ denotes the Moore-Penrose pseudoinverse.

There are several limitations of the algorithm of Elad. First, it
is suboptimal since the $t$-averaged mutual coherence $\mu_t(\M)$
is different from the mutual coherence $\mu(\M)$ which is our real
target. Second, the proposed algorithm to minimize $\mu_t(\M)$ has
no convergence guarantee. So the quality of the obtained solution is not
guaranteed. Third, the choices of two parameters, $t$ and
$\gamma$, are crucial for the signal recovery performance in CS.
However, there is no guideline for their settings and thus in
practice it is usually difficult to find their best choices.

\subsubsection{The Algorithm of Duarte-Carajalino and Sapiro}
The algorithm of Duarte-Carajalino and Sapiro
\cite{duarte2009learning} is not a method that is based on mutual
coherence. It instead aims to find the sensing matrix $\PP$ such
that the corresponding Gram matrix is as close to the identity
matrix as possible, i.e.,
\begin{equation}\label{dcs1}
\G=\M^T\M=\D^T\PP^T\PP\D\approx\I,
\end{equation}
where $\I$ denotes the identity matrix. Multiplying both sides of
the previous expression by $\D$ on the left and $\D^T$ on the
right, it becomes
\begin{equation}\label{dcs2}
\D\D^T\PP^T\PP\D\D^T\approx\D\D^T.
\end{equation}
Let $\D\D^T=\V\bm\Lambda\V^T$ be the eigen-decomposition of $\D\D^T$. Then (\ref{dcs2}) is equivalent to
\begin{equation}
\Lambda\V^T\PP^T\PP\V\Lambda=\bm\Lambda.
\end{equation}
Define $\bm\Gamma=\PP\V$. Then they finally formulate the following model w.r.t. $\bm\Gamma$
\begin{equation}
\min_\Gamma \norm{\bm\Lambda-\bm\Lambda\bm\Gamma^T\bm\Gamma\bm\Lambda}_F.
\end{equation}
After solving the above problem, the projection matrix can be
obtained as $\PP=\bm\Gamma\V^T$.

However, usually the signal recovery performance of the algorithm
of Duarte-Carajalino and Sapiro is not very good. The reason is
that $\M$ is overcomplete and the Gram matrix $\G$ cannot be an
identity matrix. In this case, simply minimizing the difference
between the Gram matrix $\G$ and the identity matrix does not
imply a solution $\M$ with low mutual coherence.

\subsubsection{The Algorithm of Xu et al.}
The algorithm of Xu et al. \cite{xu2010optimized} is motivated by
the well-known Welch bound \cite{welch1974lower}. For any
$\M\in\mathbb{R}^{m\times n}$, the mutual coherence $\mu(\M)$ is
lower bounded, e.g.,
\begin{equation}\label{welchbound}
\mu(\M) \geq \sqrt{\frac{n-m}{m(n-1)}}.
\end{equation}
The algorithm of Xu et al. aims to find $\M$ such that the
off-diagonal elements of $\G=\M^T\M$ approximate the Welch bound
well. They proposed to solve the following problem
\begin{equation}\label{xu}
\begin{split}
& \min_{\G} \norm{\G-\G_{\Lambda}}_F \\
\text{s.t.} \ \G_{\Lambda}=\G_{\Lambda}^T, \ &\text{diag}(\G_{\Lambda})=\1, \ \norm{\G_{\Lambda}}_{\infty,\text{off}}\leq \mu_{W},
\end{split}
\end{equation}
where $\mu_{W}=\sqrt{\frac{n-m}{m(n-1)}}$. The proposed iterative
solver for the above problem is similar to the algorithm of Elad.
The main difference is the shrinkage function used to control the
elements of $\G$. See \cite{xu2010optimized} for more details.

However, their proposed solver in \cite{xu2010optimized} for
(\ref{xu}) also lacks convergence guarantee. Another issue is
that, for $\M\in\mathbb{R}^{m\times n}$, the Welch bound
(\ref{welchbound}) is not tight when $n$ is large. Actually, the
equality of (\ref{welchbound}) can hold only when
$n\leq\frac{m(m+1)}{2}$. This implies that the algorithm of Xu et
al. is not optimal when $n>\frac{m(m+1)}{2}$.

\canyi{Beyond the above three methods, there are also some other mutual coherence optimization based  methods for the dictionary learning. For example, the work \cite{bao2014convergent} proposes a joint sparse coding and incoherent dictionary learning model which shares a similar idea as the algorithm of Duarte-Carajalino and Sapiro \cite{duarte2009learning}. The work \cite{barchiesi2013learning} considers a model with hard constraint on the mutual coherence and sparsity and proposes a heuristic iterative projection solver.   Greedy algorithms are proposed in \cite{li2011dictionaries,schnass2008dictionary} to find a sensing matrix for a dictionary that gives low cumulative coherence.
}

\subsection{Contributions}
There are at least two main issues in the previous methods
reviewed above. First, none of them aims to find $\PP$ by directly
minimizing $\mu(\PP\D)$ which is our real target. Thus the
objectives of these methods are not optimal. For their obtained
solutions $\PP$, $\mu(\PP\D)$ is usually much larger than the
Welch bound in (\ref{welchbound}). Second, the algorithms of Elad
and Xu et al. have no convergence guarantee and thus they may
produce very different solutions given slightly different
initializations. The convergence issue may limit their
applications in CS.

To address the above issues, we develop Direct Mutual Coherence
Minimization (DMCM) models. First, we show how to construct a low
mutual coherence matrix $\M$ by minimizing $\mu(\M)$ directly.
This leads to a nonconvex and nonsmooth problem. To solve our new
problem efficiently, we first smooth the objective function such
that its gradient is Lipschitz continuous. Then we solve the
approximate problem by proximal gradient which has convergence
guarantee. Second, inspired by DMCM, we propose a DMCM based
Projection (DMCM-P) model which aims to find a projection $\PP$ by
minimizing $\mu(\PP\D)$ directly. To solve the nonconvex DMCM-P
problem, we then propose an alternating minimization method and
prove its convergence. Experimental results show that our DMCM-P
achieves the lowest mutual coherence of $\PP\D$ and also leads to
the best signal recovery performance.

\section{Low Mutual Coherence Matrix Construction}
In this section, we show how to construct a matrix
$\M\in\mathbb{R}^{m\times n}$ with low mutual coherence $\mu(\M)$
by DMCM. Assume that each column of $\M$ is normalized to unit
Euclidean length. Then we aim to find $\M$ by the following DMCM
model
\begin{equation}
\begin{split}
&\min_{\M\in\mathbb{R}^{m\times n}} \mu(\M)=\norm{\M^T\M}_{\infty,\text{off}}\\
&\st   \ \norm{\M_i}_2=1, \ i=1,\cdots,n,
\end{split}
\end{equation}
where $\M_i$ (or $(\M)_i$) denotes the $i$-th column of $\M$. The above problem is equivalent to
\begin{equation}\label{prolcd}
\begin{split}
&\min_{\M\in\mathbb{R}^{m\times n}} f(\M)=\norm{\M^T\M-\I}_{\infty}\\
& \st   \ \norm{\M_i}_2=1, \ i=1,\cdots,n,
\end{split}
\end{equation}
where $\|\A\|_\infty=\max_{i,j}|a_{ij}|$ denotes the
$\ell_\infty$-norm of $\A$. Solving the above problem is not easy
since it is nonconvex and its objective is nonsmooth. In general,
due to the nonconvexity, the globally optimal solution to
(\ref{prolcd}) is not computable. We instead consider finding a
locally optimal solution with convergence guarantee.

First, to ease the problem, we adopt the smoothing technique in
\cite{nesterov2005smooth} to smooth the nonsmooth
$\ell_{\infty}$-norm in the objective of (\ref{prolcd}). By the
fact that the $\ell_1$-norm is the dual norm of the
$\ell_\infty$-norm, the objective function in (\ref{prolcd}) can
be rewritten as
\begin{equation*}
f(\M)=\norm{\M^T\M-\I}_{\infty}=\max_{\|\V\|_1\leq1} \ \langle\M^T\M-\I,\V\rangle,
\end{equation*}
where $\|\V\|_1=\sum_{ij}|v_{ij}|$ denotes the $\ell_1$-norm of
$\V$. Since $\{\V|\|\V\|_1\leq1\}$ is a bounded convex set, we can
define a proximal function $d(\V)$ for this set, where $d(\V)$ is
continuous and strongly convex on this set. A natural choice of
$d(\V)$ is $d(\V)=\frac{1}{2}\|\V\|_F^2$, where $\|\cdot\|_F$
denotes the Frobenius norm of a matrix. Hence, we have the
following smooth approximation of $f$ defined in (\ref{prolcd}):
\begin{equation}\label{prolcd2}
f_{\rho}(\M)=\max_{\|\V\|_1\leq1} \langle\M^T\M-\I,\V\rangle-\frac{\rho}{2}\norm{\V}_F^2,
\end{equation}
where $\rho>0$ is a smoothing parameter. Note that the smooth
function $f_{\rho}$ can approximate the nonsmooth $f$ with an
arbitrary precision and it is easier to be minimized. Indeed, $f$
and $f_\rho$ have the following relationship
\begin{equation*}
    f_\rho(\M)\leq  f(\M)\leq f_\rho(\M)+\rho\gamma,
\end{equation*}
where
$\gamma=\max_{\V}\{\frac{1}{2}\norm{\V}_F^2|\norm{\V}_{\infty}\leq1\}$.
For any $\epsilon>0$, if we choose $\rho=\frac{\epsilon}{\gamma}$,
then $|f(\M)-f_\rho(\M)|\leq\epsilon$. This implies that if $\rho$
is sufficiently small, then the difference between $f$ and
$f_\rho$ can be very small. This motives us to use $f_\rho$ to
replace $f$ in (\ref{prolcd}) and thus we have the following
relaxed problem
\begin{equation}\label{prolcdsm}
\begin{split}
&\min_{\M\in\mathbb{R}^{m\times n}} f_\rho(\M)\\
\st &  \ \norm{\M_i}_2=1, \ i=1,\cdots,n.
\end{split}
\end{equation}
As $f_\rho$ can approximate $f$ at an arbitrary precision, solving
(\ref{prolcdsm}) can still be regarded as directly minimizing the
mutual coherence. Problem (\ref{prolcdsm}) is easier to solve
since $\nabla f_{\rho}(\M)=\M(\V^*+{\V^*}^T)$, where $\V^*$ is the
optimal solution to (\ref{prolcd2}), is Lipschitz continuous. That
is, for any $\M_1,\M_2\in\mathbb{R}^{m\times n}$, there exists a
constant $L=1/\rho$ such that
\begin{equation*}
\|\nabla f_{\rho}(\M_1)-\nabla f_{\rho}(\M_2)\|_F\leq L\norm{\M_1-\M_2}_F.
\end{equation*}
With the above property, problem (\ref{prolcdsm}) can be solved by
the proximal gradient method which updates $\M$ in the $(k+1)$-th
iteration by
\begin{align}
\M_{k+1}=& \arg\min_{\M} \langle\nabla f_{\rho}(\M_k), \M-\M_k\rangle+\frac{1}{2\alpha}\norm{\M-\M_k}_F^2 \notag\\
=& \arg\min_{\M} \frac{1}{2}\norm{\M-\left(\M_k-\alpha\nabla f_{\rho}(\M_k) \right)}_F^2 \label{updateM} \\
\st &  \ \norm{\M_i}_2=1, \ i=1,\cdots,n,\notag
\end{align}
where $\alpha>0$ is the step size. To guarantee convergence, it is
required that $\alpha<\rho$. In this work, we simply set
$\alpha=0.99\rho$. The above problem has a closed form solution by
normalizing each column of $\M_k-\alpha\nabla f_{\rho}(\M_k)$,
i.e.,
\begin{equation}\label{eqcomputeMkk}
(\Mkk)_i=\frac{(\M_k-\alpha\nabla f_{\rho}(\M_k))_i}{\norm{(\M_k-\alpha\nabla f_{\rho}(\M_k))_i}_2}.
\end{equation}
To compute $\nabla f_{\rho}(\M_k)=\Mk(\Vk+{\Vk}^T)$, where $\Vk$
is optimal to (\ref{prolcd2}) when $\M=\Mk$, one has to solve
(\ref{prolcd2}) which is equivalent to the following problem
\begin{equation}\label{updateV}
\begin{split}
\Vk=&\arg\min_{\V} \frac{1}{2}\norm{\V-(\Mk^T\Mk-\I)/\rho}_F, \\
 \st & \ \norm{\V}_1\leq1.
\end{split}
\end{equation}
Solving the above problem requires computing a proximal projection
onto the $\ell_1$ ball. This can be done efficiently by the method
in \cite{DUChi2008projectl1}.

\begin{algorithm}[t]
    \caption{Solve (\ref{prolcdsm}) by Proximal Gradient algorithm.}
    \label{m1-out}
    \textbf{Initialize:} $k=0$, $\Mk\in\mathbb{R}^{m\times n}$, $\rho>0$, $\alpha=0.99\rho$, $K>0$. \\
    \textbf{Output:} $\M^*=\text{PG}(\M_k,\rho)$. \\
    \textbf{while} $k<K$ \textbf{do}
    \begin{enumerate}
        \item Compute $\Vk$ by solving (\ref{updateV});
        \item Compute $\Mkk$ by solving (\ref{updateM});
        \item $k=k+1$.
    \end{enumerate}
    \textbf{end while}\label{alg1}
\end{algorithm}

Iteratively updating $\V$ by (\ref{updateV}) and $\M$ by
(\ref{updateM}) leads to the Proximal Gradient (PG) algorithm for
solving problem (\ref{prolcdsm}). We summarize the whole procedure
of PG for (\ref{prolcdsm}) in Algorithm \ref{alg1}.
\canyi{For the convergence guarantee, PG can be proved to be convergent. But we omit its proof since  we will introduce a more general solver and provide the convergence guarantee in Section \ref{seclmcbp}.	For the per-iteration cost of Algorithm 1, there are two main parts.  For the update of $\M$ by (\ref{updateM}), we need to compute $\nabla_\rho f(\mathbf{M}_k)=\mathbf{M}_k(\mathbf{V}_k+\mathbf{M}_k^T)$   which costs $O(mn^2)$. For the update of  $\V$ by (\ref{updateV}), we need to compute $\mathbf{M}_k^T\mathbf{M}_k$ which costs $O(mn^2)$. Thus, the  per-iteration cost of Algorithm 1 is $O(m^2n+mn^2)$.}

%Define
%\begin{equation}
%h(\M)=\left\{\begin{array}{ll} 0, & \mbox{if } \norm{\M_i}_2=1, \ i=1,\cdots,n.,\\
%             \infty, & \mbox{otherwise}.
%             \end{array}\right.\label{h_define}
%\end{equation}
%We have the convergence property:
%\begin{theorem}\label{convergence_p1} \cite{Jerone2013}
%Let $\{\M_k\}$ be the sequence generated by PG in Algorithm (\ref{alg1}). Then $\{\M_k\}$ is bounded and any accumulation point $\M^*$  of $\{\M_k\}$ is a stationary point.%, i.e.,
%%\begin{eqnarray}
%%0\in  \partial \left[f_\rho(\M^*)+h(\M^*)\right].
%%\end{eqnarray}
%\end{theorem}

Though PG is guaranteed to converge, the obtained suboptimal
solution to (\ref{prolcdsm}) may be far from optimal to problem
(\ref{prolcd}) which is our original target. There are two
important factors which may affect the quality of the obtained
solution by PG. First, due to the nonconvexity of
(\ref{prolcdsm}), the solution may be sensitive to the
initialization of $\M$. Second, the smoothing parameter $\rho>0$
should be small so that the objective $f_\rho$ in (\ref{prolcdsm})
can well approximate the objective $f$ in (\ref{prolcd}). However,
if $\rho$ is directly set to a very small value, PG may decrease
the objective function value of (\ref{prolcdsm}) very slowly. This
can be easily seen from the updating of $\M$ in (\ref{updateM}),
where $\alpha < \rho$. To address the above two issues, we use a
continuation trick to find a better solution to (\ref{prolcd}) by
solving (\ref{prolcdsm}) with different initializations. Namely,
we begin with a relatively large value of $\rho$ and reduce it
gradually. For each fixed $\rho$, we solve (\ref{prolcdsm}) by PG
in Algorithm \ref{alg1} and use its solution as a new
initialization of $\M$ in PG. To achieve a better solution, we
repeat the above procedure $T$ times or until $\rho$ reaches a
predefined small value $\rho_{\min}$. We summarize the procedure
of PG with the continuation trick in Algorithm \ref{alg_cg}.

Finally, we would like to emphasize some advantages of our DMCM
model (\ref{prolcd}) and the proposed solver. A main merit of our
model (\ref{prolcd}) is that it minimizes the mutual coherence
$\mu(\M)$ directly and thus the mutual coherence of its optimal
solution can be low. Though the optimal solution is in general not
computable due to the nonconvexity of (\ref{prolcd}), our proposed
solver, which first smooths the objective and then minimizes it by
PG, has convergence guarantee. To the best of our knowledge, this
is the first work which directly minimizes the mutual coherence of
a matrix with convergence guarantee.

\begin{algorithm}[t]
    \caption{Solve (\ref{prolcdsm}) by PG with continuation trick.}
    \label{m1-in}
    \textbf{Initialize:} $\rho>0$, $\alpha=0.99\rho$, $\eta>1$, $\M$, $t=0$, $T>0$. \\
    \textbf{while} $t<T$ \textbf{do}
    \begin{enumerate}
        \item $\M=\text{PG}(\M,\rho)$ by calling Algorithm \ref{alg1};
        \item $\rho=\rho/\eta$, $\alpha=0.99\rho$;
        \item $t=t+1$.
    \end{enumerate}
    \textbf{end while}\label{alg_cg}
\end{algorithm}

\section{Low Mutual Coherence Based Projection}
\label{seclmcbp}
In this section, we show how to find a projection matrix $\PP$
such that $\mu(\PP\D)$ can be as low as possible. This is crucial
for signal recovery by CS associated to problem (\ref{eq_ed}).
Similar to the DMCM model shown in (\ref{prolcd}), an ideal way is
to minimize $\mu(\PP\D)$ directly, i.e.,
\begin{equation}\label{lmcpideal}
\begin{split}
&\min_{\PP\in\mathbb{R}^{m\times d}} \norm{(\PP\D)^T(\PP\D)-\I}_{\infty}\\
& \st  \ \norm{\PP\D_i}_2=1, \ i=1,\cdots,n.
\end{split}
\end{equation}
However, the constraint of (\ref{lmcpideal}) is more complex than
the one in (\ref{prolcd}), and thus (\ref{lmcpideal}) is much more
challenging to solve. We instead consider an approximate model of
(\ref{lmcpideal}) based on the following observation.
\begin{theorem}
    For any $\M_1, \M_2\in\mathbb{R}^{m\times n}$, if $\M_1\rightarrow\M_2$, then $\mu(\M_1)\rightarrow\mu(\M_2)$.
\end{theorem}
It is easy to prove the above result by the definition of the
mutual coherence of a matrix. The above theorem indicates that the
difference of the mutual coherences of two matrices is small when
the difference of two matrices is small. This motivates us to find
$\M$ such that $\mu(\M)$ is low and the difference between $\M$
and $\PP\D$ is small. So we have the following approximate model
of (\ref{lmcpideal}):
\begin{equation}
\begin{split}
    \min_{\PP\in \mathbb R^{m\times d},\M\in \mathbb R^{m\times n}}  & \ \|\M^T\M-\I\|_{\infty}+\frac{1}{2\beta}\|\M-\PP\D\|_F^2\label{prob2}\\
 \st & \ \|\M_i\|_2=1,i=1,\cdots,n,
    \end{split}
\end{equation}
where $\beta>0$ trades off $\mu(\M)$ and the difference between
$\M$ and $\PP\D$. To distinguish from the DMCM model in
(\ref{prolcd}), in this paper we name the above model as DMCM
based Projection (DMCM-P).

Now we show how to solve (\ref{prob2}). First, we smooth
$\|\M^T\M-\I\|_{\infty}$ as $f_{\rho}(\M)$ defined in
(\ref{prolcd2}). Then problem (\ref{prob2}) can be approximated by
the following problem with a smooth objective:
\begin{equation}
\begin{split}
\min_{\PP,\M}\ & F(\M,\PP)=f_{\rho}(\M)+\frac{1}{2\beta}\|\M-\PP\D\|_F^2\label{prolcdsm2}\\
\st & \ \|\M_i\|_2=1,i=1,\cdots,n.
\end{split}
\end{equation}
When both $\rho$ and $\beta$ are small, $f_{\rho}$ is very close
to $f$. So is $\mu(\PP\D)$ to $\mu(\M)$ because $\|\M-\PP\D\|_F$
has to be small. Thus solving problem (\ref{prolcdsm2}) can still
be regarded as minimizing the mutual coherence directly. We
propose to alternately update $\PP$ and $\M$ to solve problem
(\ref{prolcdsm2}).

1. Fix $\PP=\PP_k$ and update $\M$ by
\begin{align}
&\M_{k+1} \notag\\
=& \arg\min_{\M} \langle\nabla f_{\rho}(\M_k), \M-\M_k\rangle+\frac{1}{2\alpha}\norm{\M-\M_k}_F^2 \notag\\
&+\frac{1}{2\beta}\|\M-\PP_k\D\|_F^2 \label{updateM2}\\
=& \arg\min_{\M} \frac{1}{2}\norm{\M-\frac{\left(\frac{1}{\alpha}\M_k+\frac{1}{\beta}\PP_k\D-\nabla f_{\rho}(\M_k) \right)}{\frac{1}{\alpha}+\frac{1}{\beta}}}_F^2 \notag \\
& \st   \ \norm{\M_i}_2=1, \ i=1,\cdots,n,\notag
\end{align}
where $\alpha>0$ is a step size satisfying $\alpha<\rho$. Similar
to (\ref{updateM}), the above problem has a closed form solution.
To compute $\nabla f_{\rho}(\Mk)$ in (\ref{updateM2}), we also
need to compute $\Vk$ by solving (\ref{updateV}).

\begin{algorithm}[t]
    \caption{Solve (\ref{prolcdsm2}) by Alternating Minimization.}
    \label{m2-out}
    \textbf{Initialize:} $k=0$, $\PP_k\in\mathbb{R}^{m\times d}$, $\M_k\in\mathbb{R}^{m\times n}$, $\rho>0$, $\alpha=0.99\rho$, $\beta>0$.\\
    \textbf{Output:} $\{\PP^*\M^*\}=\text{AM}(\M_k,\PP_k,\rho,\beta)$. \\
    \textbf{while} $k<K$ \textbf{do}
    \begin{enumerate}
        \item Compute $\Vk$ by solving (\ref{updateV});
        \item Compute $\Mkk$ by solving (\ref{updateM2});
        \item Compute $\PP_{k+1}$ by solving (\ref{P_step});
        \item $k=k+1$.
    \end{enumerate}
    \textbf{end while}\label{altmin}
\end{algorithm}

2. Fix $\M=\Mkk$ and update $\PP$ by solving
\begin{eqnarray}
&&\PP_{k+1}=\argmin_{\PP} \|\Mkk-\PP\D\|_F^2,\label{P_step}
\end{eqnarray}
which has a closed form solution $\PP=\Mkk\D^\dagger$.

Iteratively updating $\PP$ by (\ref{P_step}) and $\M$ by
(\ref{updateM2}) leads to the Alternating Minimization (AM) method
for (\ref{prolcdsm2}). We summarize the whole procedure of AM in
Algorithm \ref{altmin}. It can be easily  seen that the
per-iteration cost of Algorithm \ref{altmin} is $O((d+m)n^2+n^3)$.
We can prove that the sequence generated by AM converges to a critical point.

We define
\begin{equation}\label{eqdefh}
h(\M)=\left\{\begin{array}{ll} 0, & \mbox{if } \|\M_i\|_2=1,i=1,\cdots,n,\\
+\infty, & \mbox{otherwise}.
\end{array}\right.
\end{equation}
\begin{CanyiPar}{black}
\begin{theorem}\label{convergence_p2}
Assume that $\D$ in problem (\ref{prolcdsm2}) is of full row rank.
Let $\{(\M_k,\PP_k)\}$ be the sequence generated by Algorithm
\ref{altmin}. Then the following results hold:
\begin{enumerate}[(i)]
    \item There esits some constants $a>0$ and $b>0$ such that
    \begin{align}
    &h(\M_{k+1})+F(\M_{k+1},\PP_{k+1})\notag\\
    \leq& h(\M_k)+F(\M_k,\PP_k)-a\|\M_{k+1}-\M_k\|_F^2- b\norm{\PP_{k+1}-\PP_{k}}_F^2.
    \end{align}
    \item There exists $\W_{k+1}\in \nabla_{\M} F(\M_{k+1},\PP_{k+1})+\partial h(\Mkk)$ and constants $c>0$, $d>0$, such that
    \begin{align}
    \norm{\W_{k+1}}_F \leq c\norm{\Mkk - \Mk}_F + d\norm{\PP_{k} - \PP_{k+1}}_F,
    \end{align}
    \begin{eqnarray}
    \nabla_{\PP} F(\M_{k+1},\PP_{k+1})=\0.
    \end{eqnarray}
    \item There exist a subsequence $\{(\M_{k_j},\PP_{k_j})\}$ and $(\M^*,\PP^*)$ such that  $(\M_{k_j},\PP_{k_j})\rightarrow(\M^*,\PP^*)$ and
    $F(\M_{k_j},\PP_{k_j})+h(\M_{k_j})\rightarrow
    F(\M^*,\PP^*)+h(\M^*)$.
\end{enumerate}
\end{theorem}
The proof of Theorem \ref{convergence_p2} can be found in
Appendix. Note that to guarantee the convergence of Algorithm
\ref{altmin}, Theorem \ref{convergence_p2} requires $\D$ in
problem (\ref{prolcdsm2}) to be of full row rank. Such an
assumption usually holds in CS since $\D\in\mathbb{R}^{d\times n}$
is an overcomplete dictionary with $d<n$.

Based on Theorem \ref{convergence_p2}, we then have the following convergence results.
\begin{theorem}\label{thm5concirpoint}
	 (Convergence to a critical point). The sequence  $\{(\M_k,\PP_k)\}$   generated by Algorithm
	 \ref{altmin} converges to a critical point of $F(\M,\PP)+h(\M)$. Moreover, the sequence  $\{(\M_k,\PP_k)\}$ ha a finite length, i.e.,
	 \begin{equation*}
	 \sum_{k=0}^{+\infty} \left( a \norm{\Mkk-\Mk} + b\norm{\PP_{k+1}-\PP_k}\right) < \infty,
	 \end{equation*}
	 where $a>0$ and $b>0$ are constants as in Theorem  \ref{convergence_p2} (i).
\end{theorem}

Theorem \ref{thm5concirpoint} is directly obtained by Theorem 2.9 in \cite{attouch2013convergence}  based on the results in Threorem \ref{convergence_p2}.
\end{CanyiPar}
Though AM is guaranteed to converge, the obtained solution to
(\ref{prolcdsm2}) may be far from optimal to problem (\ref{prob2})
which is our original target. In order for (\ref{prolcdsm2}) to
approximate (\ref{prob2}) well, $\rho>0$ should be small. On the
other hand, $\beta>0$ should also to be small such that the
difference between $\M$ and $\PP\D$ is small and thus $\mu(\PP\D)$
can well approximate $\mu(\M)$. Similar to Algorithm \ref{alg_cg},
we use a continuation trick to achieve a good solution to
(\ref{prob2}).
%
%
% However, if $\rho$ and $\beta$ are directly set to very small values, AM will decrease the objective function value of (\ref{prolcdsm2}) slowly. To balance the convergence speed and the quality of the obtained solution, we use a continuation trick as that in Algorithm \ref{alg_cg}.
Namely, we begin with a relatively large value of $\rho>0$ and
$\beta>0$ and reduce them gradually. For each fixed pair
$(\rho,\beta)$, we solve (\ref{prolcdsm2}) by AM in Algorithm
\ref{altmin} and use its solution as a new initialization of $\PP$
and $\M$ in AM. We repeat the procedure $T$ times or until $\rho$
and $\beta$ reach predefined small values $\rho_{\min}$ and
$\beta_{\min}$. We summarize the procedure of AM with the
continuation trick in Algorithm \ref{continution_altmin}.

Finally, we would like to emphasize some advantages of our DMCM-P
over previous methods. The main merit of our DMCM-P is that it is
the first model which minimizes $\mu(\PP\D)$ directly and the
proposed solver also has convergence guarantee. The algorithms of
Elad \cite{elad2007optimized} and Xu et al. \cite{xu2010optimized}
are also mutual coherence based methods. But their objectives are
suboptimal and their solvers lack convergence guarantee.

% Compared with the algorithm of Elad which aims to minimize the $t$-averaged mutual coherence $\mu_t(\M)$, our DMCM-P can minimize $\mu(\M)$ directly. Note that our optimization solver in Algorithm \ref{continution_altmin} guarantees that $\|\M-\PP\D\|_F$ can be arbitrarily small, and thus the $\mu(\PP\D)$ is expected to be as small as $\mu(\M)$ (give a figure to show this).

\canyi{It is worth mentioning that the sparse signal recovery can be guaranteed under some other different settings and   conditions. The low mutual coherence   property still plays an important role.  For example, a similar recovery bound can be obtained under the additional assumption that the signs of the non-zero entries of the signal are chosen at random  \cite{rauhut2010compressive,candes2007sparsity}. The theory requires incoherence
	between the sensing and sparsity bases. The variable density sampling is a technique to recover the signal of highest sparsity by optimizing the   sampling profile \cite{puy2011variable}. The proposed technique which directly minimizes the mutual coherence may be also applied in  the variable density sampling to improve the recovery performance.
}

\begin{algorithm}[t]
\caption{Solve (\ref{prolcdsm2}) by AM with continuation trick.}
\label{m2-in}
\textbf{Initialize:} $\rho>0$, $\alpha=0.99\rho$, $\beta>0$, $\eta>1$, $\M$, $\PP$,  $t=0$, $T>0$. \\
    \textbf{while} $t<T$ \textbf{do}
    \begin{enumerate}
        \item $(\PP,\M)=\text{AM}(\PP,\M,\rho,\beta)$ by calling Algorithm \ref{altmin};
        \item $\rho=\rho/\eta $, $\alpha=0.99\rho$;
        \item $\beta=\beta/\eta$;
        \item $t=t+1$.
    \end{enumerate}
\textbf{end while}\label{continution_altmin}
\end{algorithm}

\begin{figure*}
    \begin{subfigure}[b]{0.32\textwidth}
        \centering
        \includegraphics[width=\textwidth]{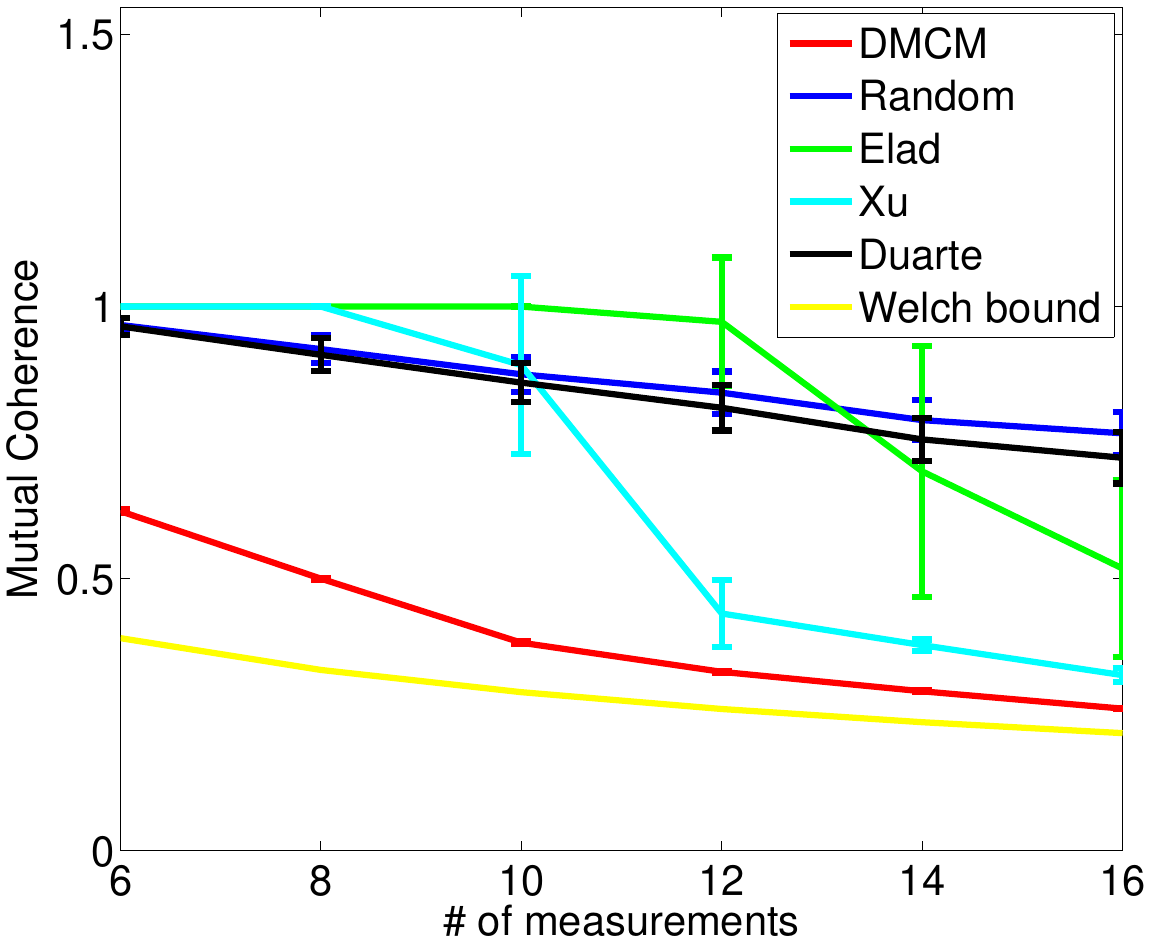}
        \caption{$n=60$}
    \end{subfigure}
    \begin{subfigure}[b]{0.32\textwidth}
        \centering
        \includegraphics[width=\textwidth]{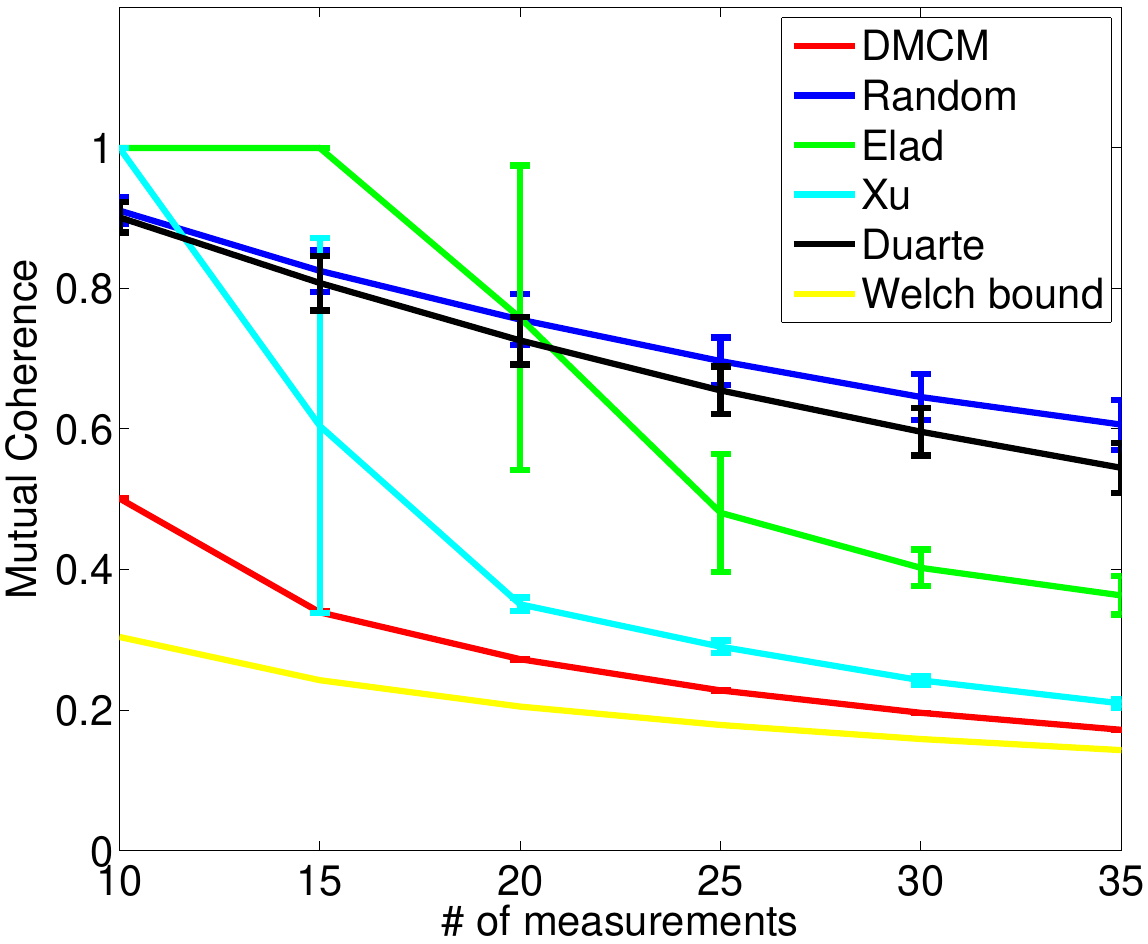}
        \caption{$n=120$}
    \end{subfigure}
    \begin{subfigure}[b]{0.32\textwidth}
        \centering
        \includegraphics[width=\textwidth]{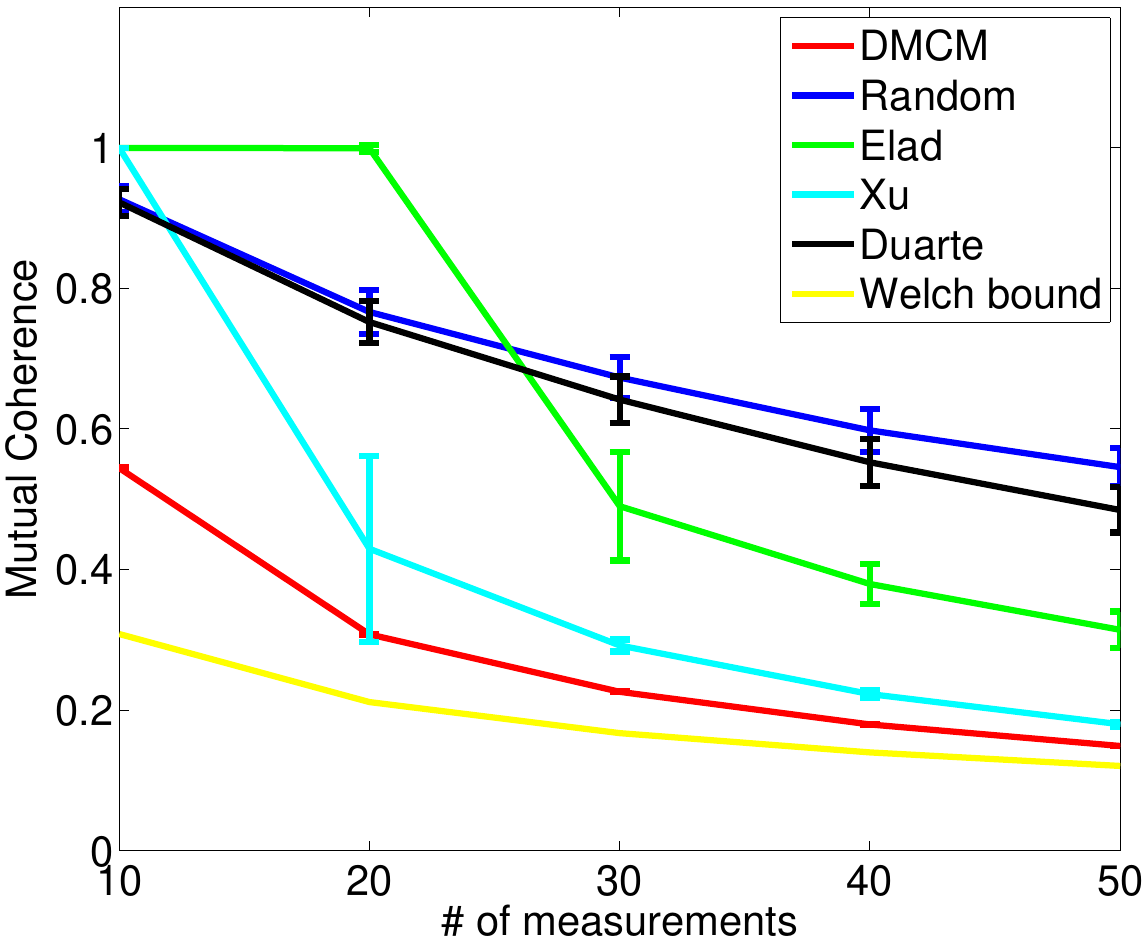}
        \caption{$n=180$}
    \end{subfigure}
    \caption{Plots of the means and standard deviations of mutual coherences of $\M$ v.s. the number $m$ of measurements.}\label{fig_mcvsnummeas}
    %\vspace{-1em}
\end{figure*}
 \begin{figure*}
    \begin{subfigure}[b]{0.32\textwidth}
        \centering
        \includegraphics[width=\textwidth]{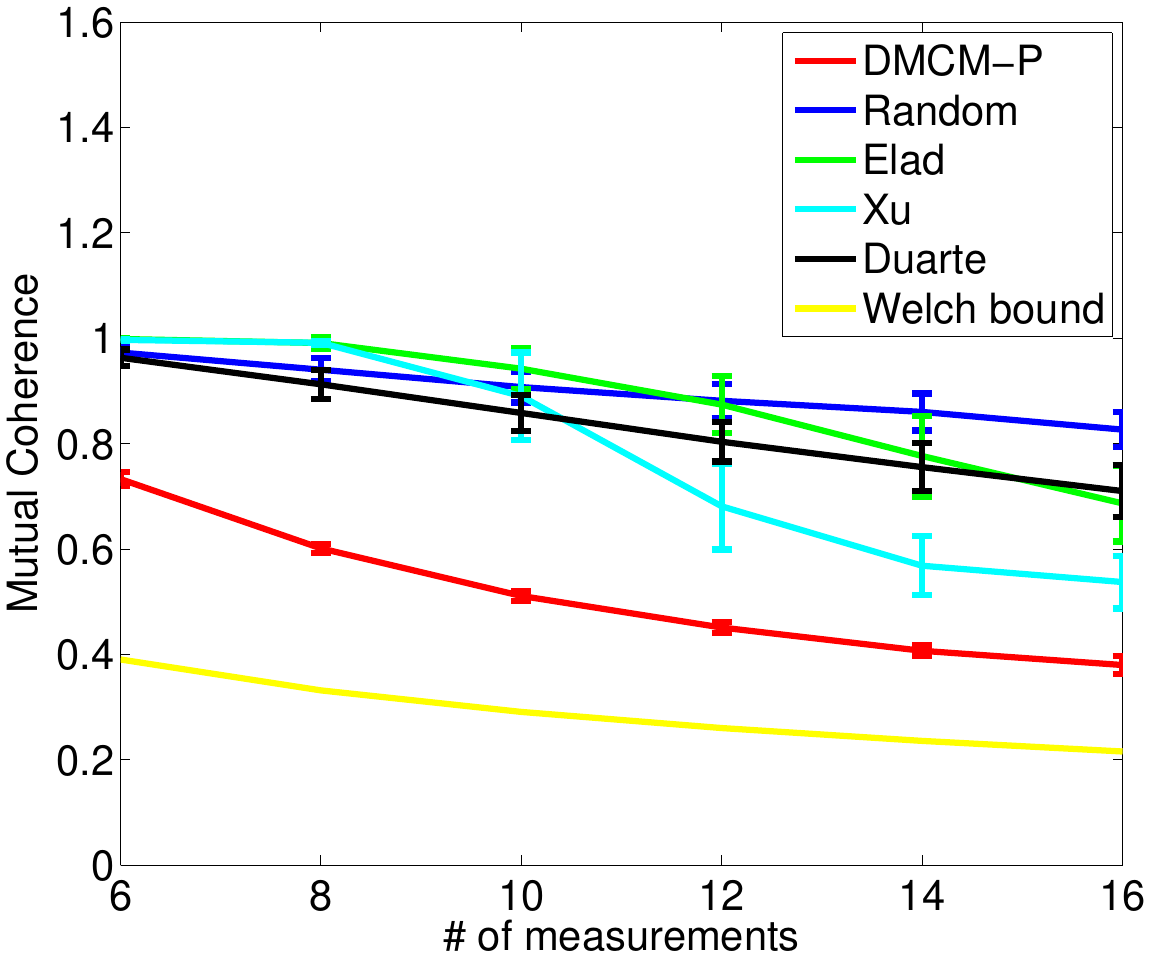}
        \caption{$n=60$}
    \end{subfigure}
    \begin{subfigure}[b]{0.32\textwidth}
        \centering
        \includegraphics[width=\textwidth]{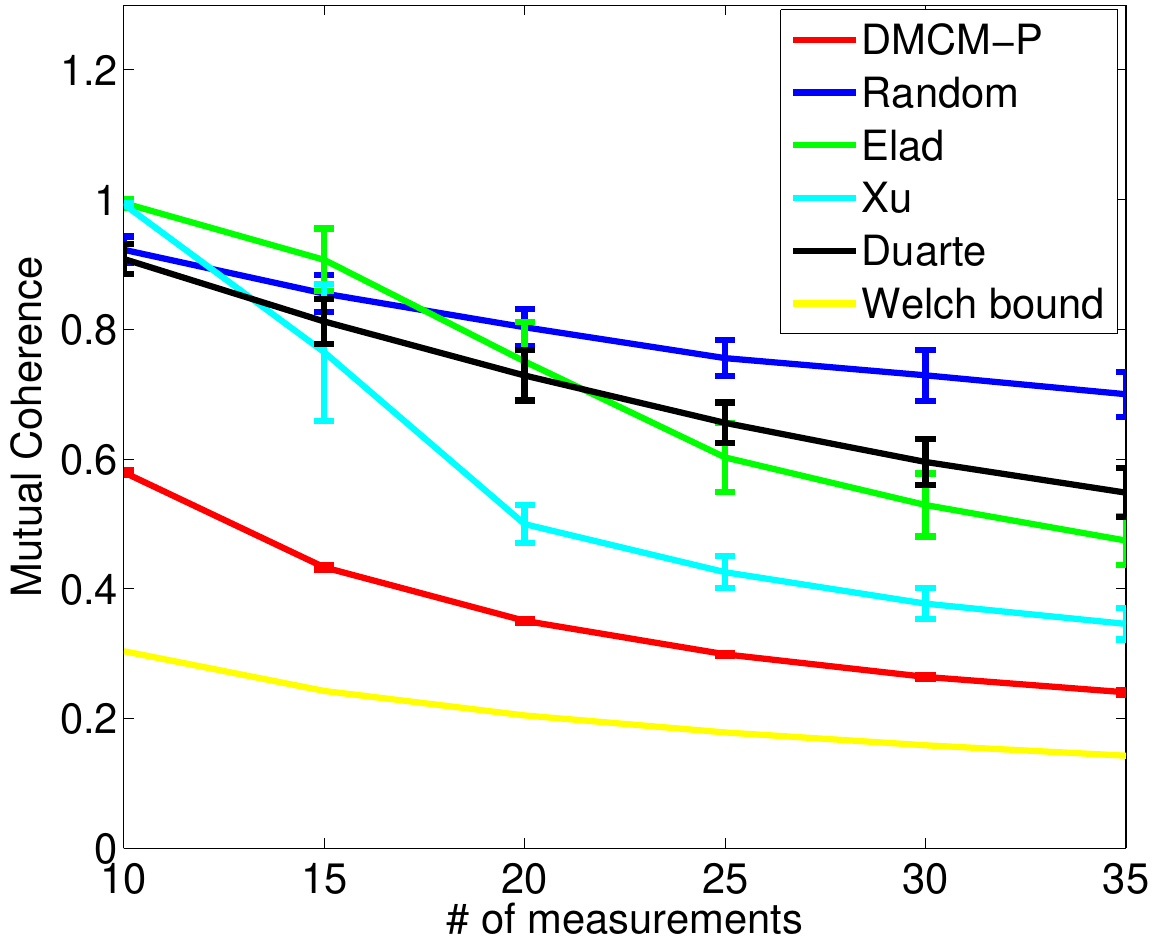}
        \caption{$n=120$}
    \end{subfigure}
    \begin{subfigure}[b]{0.32\textwidth}
        \centering
        \includegraphics[width=\textwidth]{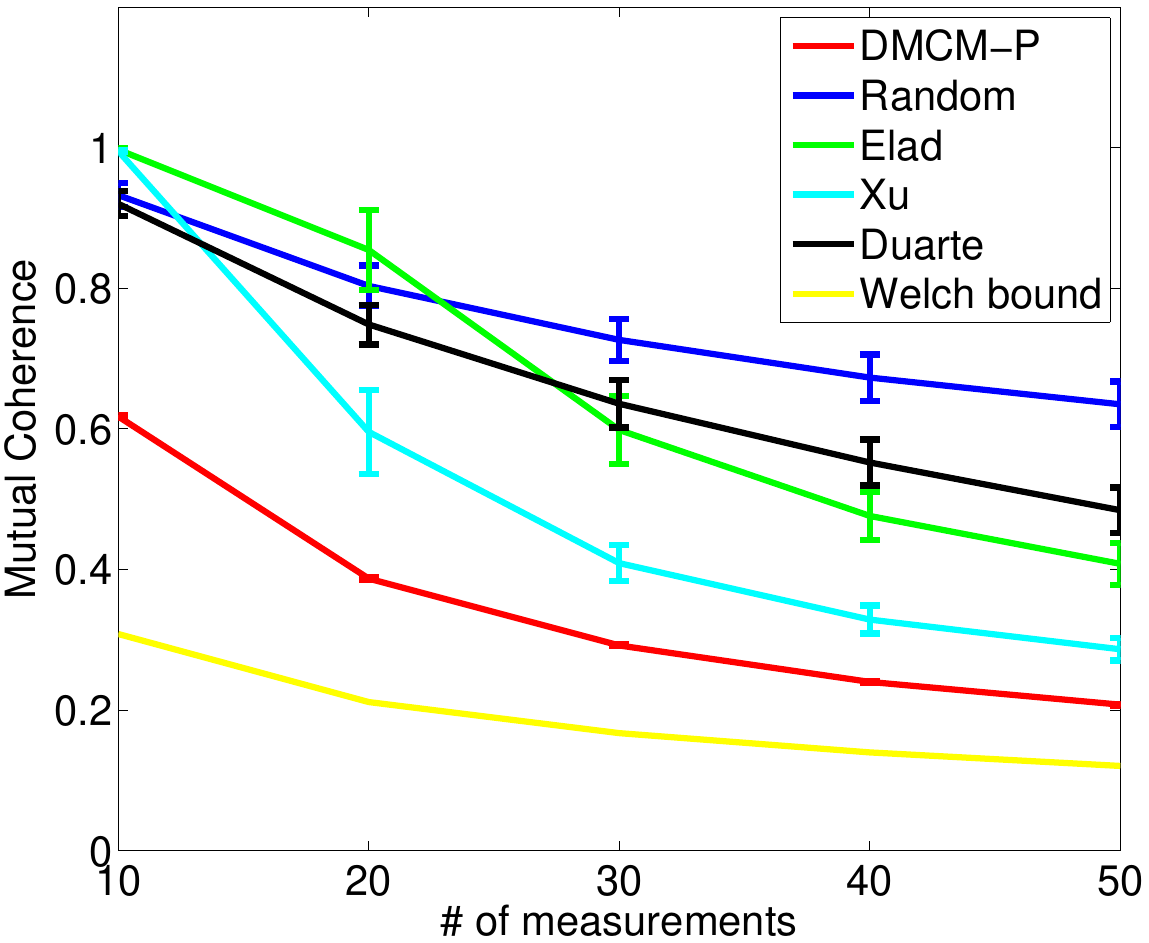}
        \caption{$n=180$}
    \end{subfigure}
    \caption{Plots of the means and standard deviations of mutual coherences of $\PP\D$ v.s. the number $m$ of measurements, where $\D$ is a standard Gaussian random matrix.}\label{fig_mcvsproj}
    %\vspace{-1em}
 \end{figure*}
\section{Numerical Results}
In this section, we conduct several experiments to verify the
effectiveness of our proposed methods by comparing them with
previous methods. The experiments consist of two parts. The first
part shows the values of mutual coherence. The second part shows
the signal recovery errors in CS.

\subsection{Comparing the Mutual Coherence}

This subsection presents two experiments to show the effectiveness
of DMCM and DMCM-P, respectively. In the first experiment, we show
that our DMCM is able to construct a matrix
$\M\in\mathbb{R}^{m\times n}$ with lower mutual coherence than
previous methods do. We compare DMCM with
\begin{itemize}
    \item Random: random matrix whose elements are drawn independently from the standard normal distribution.
    \item Elad: the algorithm of Elad \cite{elad2007optimized} with $\D=\I$.
    \item Xu: the algorithm of Xu et al. \cite{xu2010optimized} with $\D=\I$.
    \item Duarte: the algorithm of Duarte-Carajalino and Sapiro \cite{duarte2009learning} with $\D=\I$.
    \item Welch bound: the Welch bound \cite{welch1974lower} shown in (\ref{welchbound}).
\end{itemize}
Note that the compared algorithms of Elad
\cite{elad2007optimized}, Xu et al. \cite{xu2010optimized} and
Duarte-Carajalino and Sapiro \cite{duarte2009learning} were
designed to find a projection $\PP$ such that $\M=\PP\D$ has low
mutual coherence. They can still be compared with our DMCM by
setting $\D$ as the identity matrix $\I$.

To solve our DMCM model in (\ref{prolcdsm}), we run Algorithm
\ref{m1-in} for 15 iterations and Algorithm \ref{m1-out} for 1000
iterations. In Algorithm \ref{alg_cg}, we set $\rho_0=0.5$ and
$\eta=1.2$. $\M$ is initialized as a Gaussian random matrix. In
the method of Elad, we follow \cite{elad2007optimized} to set
$t=0.2$ and $\gamma = 0.95$. In the method of Xu, we try multiple
choices of the convex combination parameter $\alpha$ and set it as
0.5 which results in the lowest mutual
coherence in most cases. The method of Duarte do not need special parameters. % All the compared methods have the same initializations.
All the compared methods have the same random initializations of
$\PP$ (except Duarte, which has a closed form solution).

The compared methods are tested on three settings with different
sizes of $\M\in\mathbb{R}^{m\times n}$: (1) $m=[6:2:16],n=60$; (2)
$m=[10:5:35],n=120$; and (3) $m=[10:10:50],n=180$. Note that the
constructed matrices may not be the same for the compared methods
with different initializations. So for each choice of size
$(m,n)$, we repeat the experiment for 100 times and record the
means and standard deviations of the mutual coherences of the
constructed matrices $\M$. The means and standard deviations of
mutual coherences v.s. the number $m$ of measurements are shown in
Figure \ref{fig_mcvsnummeas}. It can be seen that the matrix
constructed by our DMCM achieves much lower mutual coherences than
previous methods do. The main reason is that our DMCM minimizes
the mutual coherence of $\M$ directly, while the objectives of all
the previous methods are indirect. It can also be seen that the
standard deviations of our method is close to zero, while some
other compared methods may not be stable in some cases. A possible
reason is that the solver of our method has convergence guarantee,
while other methods do not.

\begin{figure}[t]
    \centering
    \includegraphics[width=0.48\textwidth]{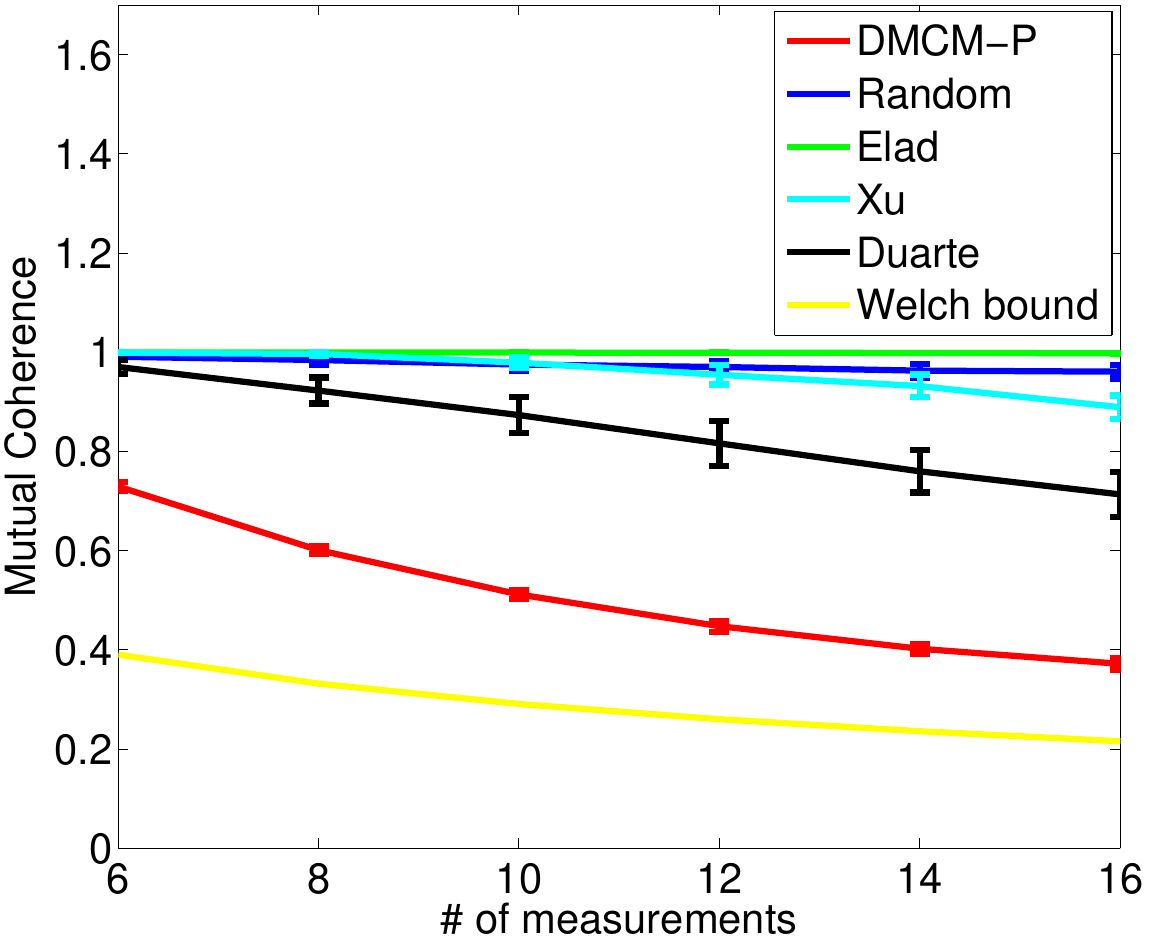}
    \caption{Plots of the means and standard deviations of mutual coherences of $\PP\D$ v.s. the number $m$ of measurements, where the elements of $\D$ are uniformly distributed in $[0, 1]$.}
    \label{mc_uniform_dist}
\end{figure}
\begin{table}[!t]
    \caption{Comparison of running time (in seconds) of DMCM-P, Elad, Xu and Duarte on problem (\ref{prob2}) under different settings.}\label{tabtime}
    \begin{center}
        \begin{tabular}{c|c|c|c|c}
            \hline
              & DMCM-P & Elad & Xu & Duarte \\
            \hline
            \hline
            $m=10$, $d=30$, $n=60$    &  181    & 5    & 5    & 0.0033   \\
            $m=20$, $d=60$, $n=120$    &  582    & 8    & 8    & 0.004   \\
            {\color{black}$m=30$, $d=90$, $n=180$}    &  838    & 14    & 12    & 0.004   \\
            \hline
        \end{tabular}
    \end{center}
\end{table}

For the second experiment in this subsection, we show that for
given $\D\in\mathbb{R}^{d\times n}$ our DMCM-P is able to compute
a projection $\PP\in\mathbb{R}^{m\times d}$ such that
$\PP\D\in\mathbb{R}^{m\times n}$ has low mutual coherence. We
choose $\D$ to be a Gaussian random matrix in this experiment. To
solve our DMCM-P model in (\ref{prob2}), we run Algorithm
\ref{m2-in} for 15 iterations and Algorithm \ref{m2-out} for 1000
iterations. In Algorithm \ref{m2-in}, we set $\rho_0=0.5$,
$\beta=2$ and $\eta=1.2$. $\PP$ is initialized as a Gaussian
random matrix.

We compare our DMCM-P with the algorithms of Elad
\cite{elad2007optimized}, Xu et al. \cite{xu2010optimized} and
Duarte-Carajalino and Sapiro \cite{duarte2009learning} on the
mutual coherence of $\PP\D$. We test on three settings: (1)
$m=[6:2:16]$, $n=60$, $d=30$; (2) $m=[10:5:35]$, $n=120$, $d=60$;
and (3) $m=[10:10:50]$, $n=180$, $d=90$. Figure \ref{fig_mcvsproj}
shows the mutual coherence of $\PP\D$ as a function of the number
$m$ of measurements. It can be seen that our DMCM-P achieves the
best projection such that $\PP\D$ has the lowest mutual coherences
in all the three settings. So are the standard deviations. Note
that our algorithm does not use any special property of $\D$. So
it is expected to work for $\D$ in other distributions as well. We
test our method in the case that the elements of $\D$ are
uniformly distributed in $[0,1]$ and report the results in Figure
\ref{mc_uniform_dist}. It can be seen that our method still
outperforms other methods in both mean and standard deviation.

\begin{figure}[t]
    \centering
    \includegraphics[width=0.48\textwidth]{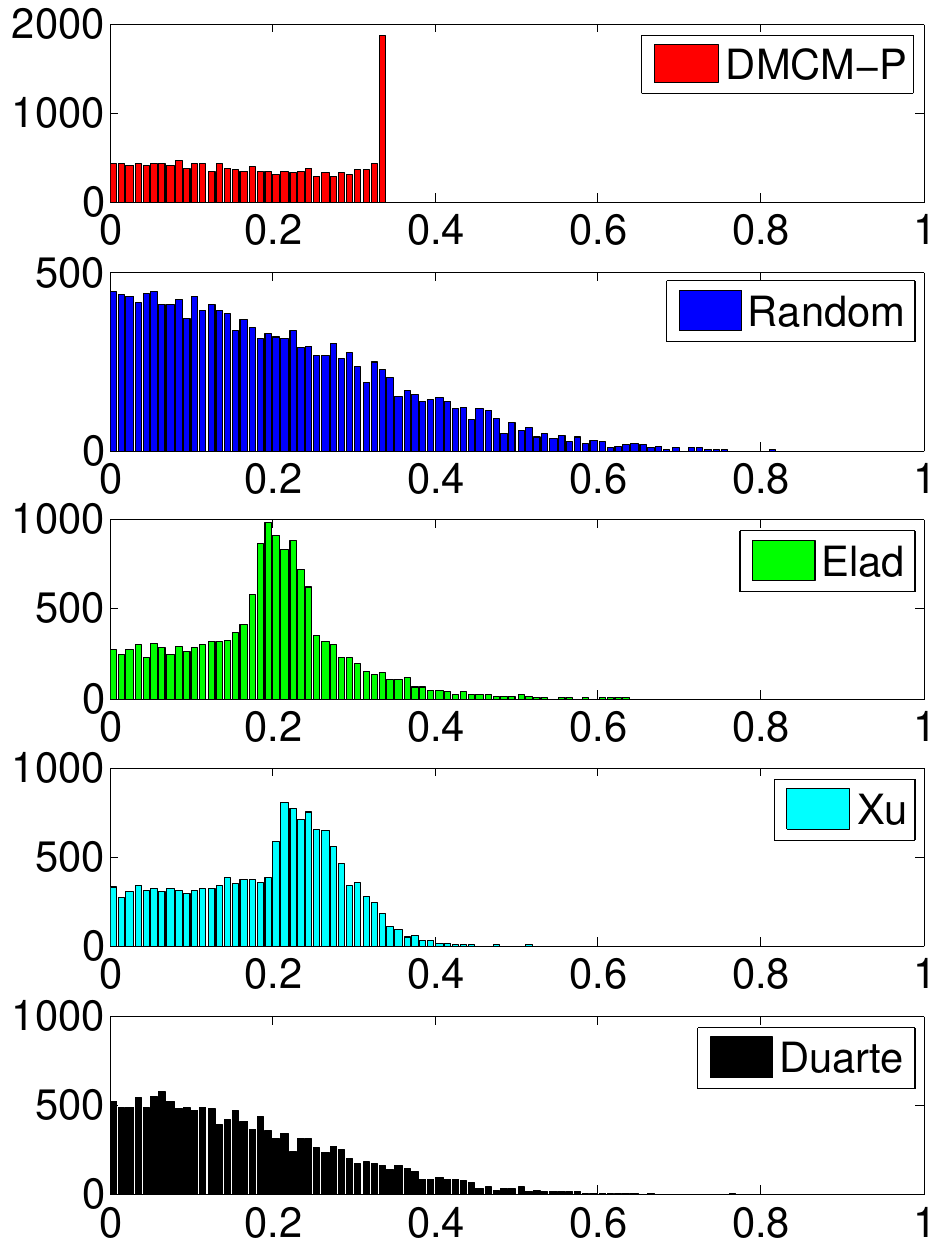}
    \caption{Distributions of the absolute values of $(\PP\D)^T(\PP\D)$.}
    \label{fig_3_errorvssparsity}
\end{figure}

Furthermore, Figure \ref{fig_3_errorvssparsity} shows the
distribution of the absolute values of inner products between
distinct columns of $\PP\D$ with $m=20$, $n=120$, and $d=60$. It
can be seen that our DMCM-P has the shortest tail, showing that
the number of elements in the Gram matrix that are closer to the
ideal Welch bound is larger than the compared methods. Such a
result is consistent with the lowest mutual coherences shown in
Figure \ref{fig_mcvsproj}.

Finally, we report the running time of the algorithms of Elad, Xu,
Duarte and our DMCM-P in Table \ref{tabtime}. The settings of the
algorithms are the same as those in Figure \ref{fig_mcvsproj} and
the running time is reported based on different choices of $m$,
$d$ and $n$. It can be seen that Duarte is the fastest method
since it has a closed form solution. Our DMCM-P is not very
efficient since we use the continuation trick in Algorithm
\ref{continution_altmin}, which repeats Algorithm \ref{altmin} many times. Note
that speeding up the algorithm, although valuable, is not the main
focus of this paper. Actually, for many applications the
projection matrix $\PP$ can be computed offline. So we leave the
speeding-up issue as future work.

\subsection{Comparing the CS Performance}

In this subsection, we apply the optimized projection by our
DMCM-P to CS. We first generate a $T$-sparse vector $\bm\alpha\in
\mathbb{R}^{n}$, which constitutes a sparse representation of
signal $\x=\D\bm\alpha$, where $\x\in\mathbb{R}^{d}$. The
locations of nonzeros are chosen randomly and their values obey a
uniform distribution in $[-1,1]$. We choose the dictionary
$\D\in\mathbb{R}^{d\times n}$ as a Gaussian random matrix, {\color{black} the DCT matrix and the matrix learned by K-SVD, respectively}. Then we
apply different projection matrices $\PP$ learned by our DMCM-P,
random projection matrix, and the algorithms of Elad
\cite{elad2007optimized}, Xu et al. \cite{xu2010optimized} and
Duarte-Carajalino and Sapiro \cite{duarte2009learning} to generate
the compressed $\y$ via $\y=\PP\D\bm\alpha$. At last, we solve
problem (\ref{eq_ed}) by OMP to obtain $\hat{\bm{\alpha}}$. We
compare the performance of projection matrices computed by
different methods using the relative reconstruction error
$\|\x-\x^*\|_2/\|\x^*\|_2$ {\color{black}and the support recovery rate $|\mbox{support}(\x)\cap \mbox{support}(\x^*)|/|\mbox{support}(\x^*)|$}, where $\x^*$ is the ground truth. A
smaller reconstruction error {\color{black} and larger support recovery rate} mean better CS performance.

We conduct two experiments in this subsection. The first one
changes the number $m$ of measurements and the second one changes
the sparsity level $T$. For every value of the aforementioned
parameters we perform 3000 experiments and calculate the average
relative reconstruction error {\color{black} and support recovery rate}.

\begin{figure}
    \centering
    \begin{tabular}{c@{\extracolsep{0.5em}}c@{\extracolsep{0.5em}}c@{\extracolsep{0.5em}}}
    \includegraphics[width=0.48\textwidth]{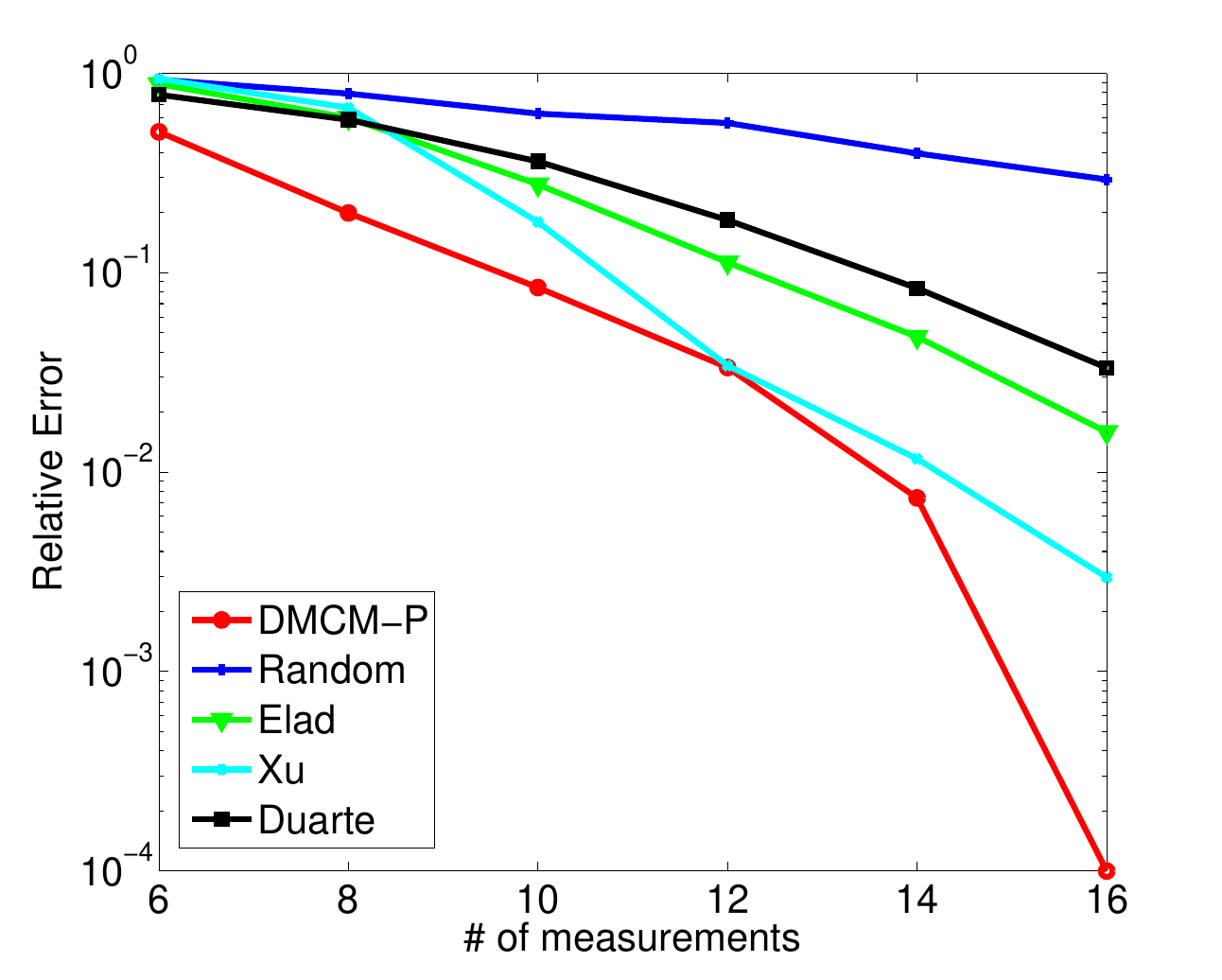}
    &\includegraphics[width=0.53\textwidth]{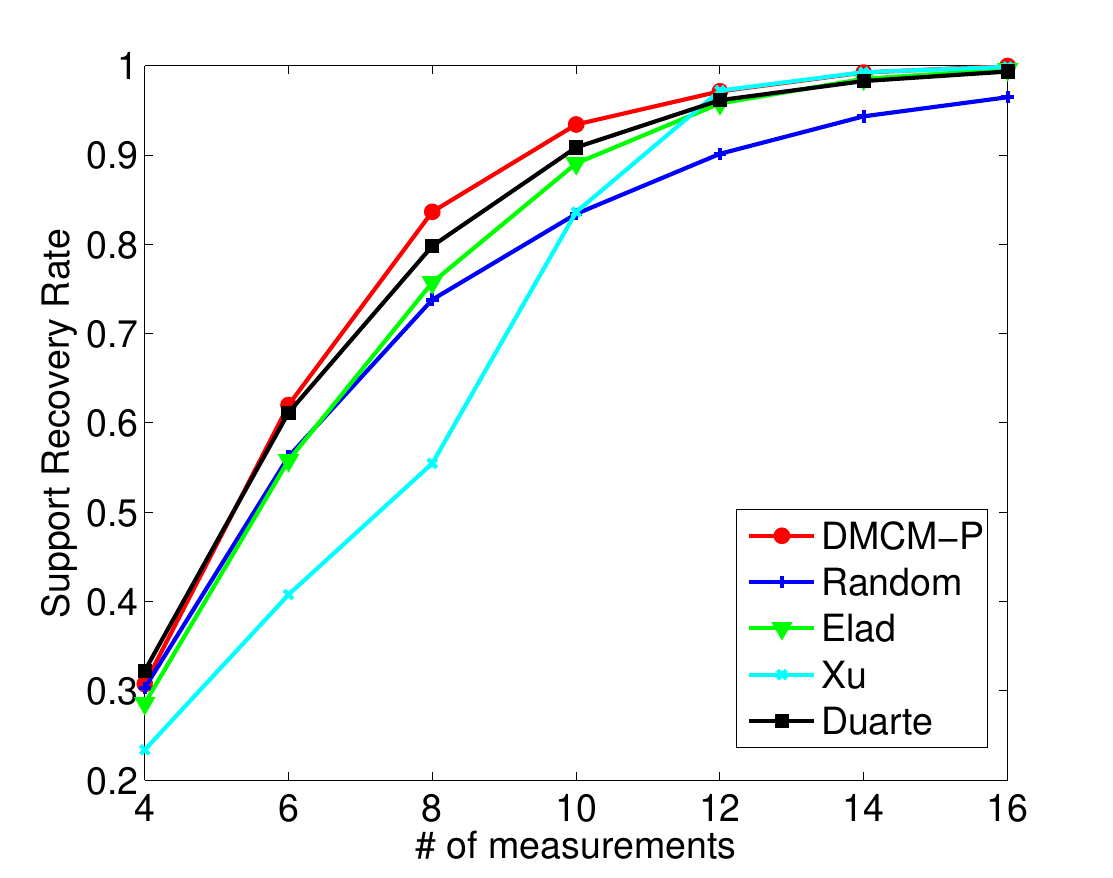}
    \end{tabular}
    \caption{Signal reconstruction errors and support recovery rate v.s. number of measurements, where $\D$ is the Gaussian random matrix.}
    \label{fig_cs_gaussian_measurement}
\end{figure}
\begin{figure}
    \centering
    \begin{tabular}{c@{\extracolsep{0.5em}}c@{\extracolsep{0.5em}}c@{\extracolsep{0.5em}}}
    \includegraphics[width=0.53\textwidth]{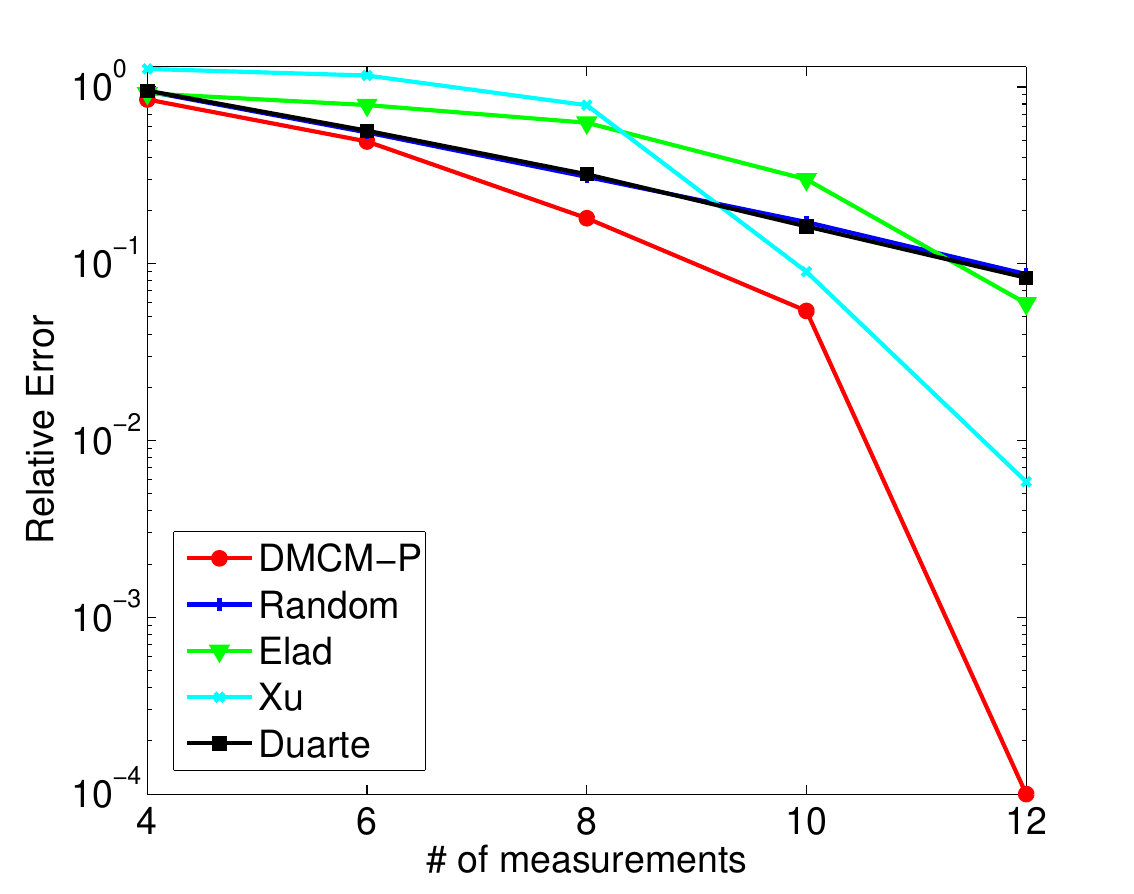}
    &\hspace{-0.6cm}\includegraphics[width=0.53\textwidth]{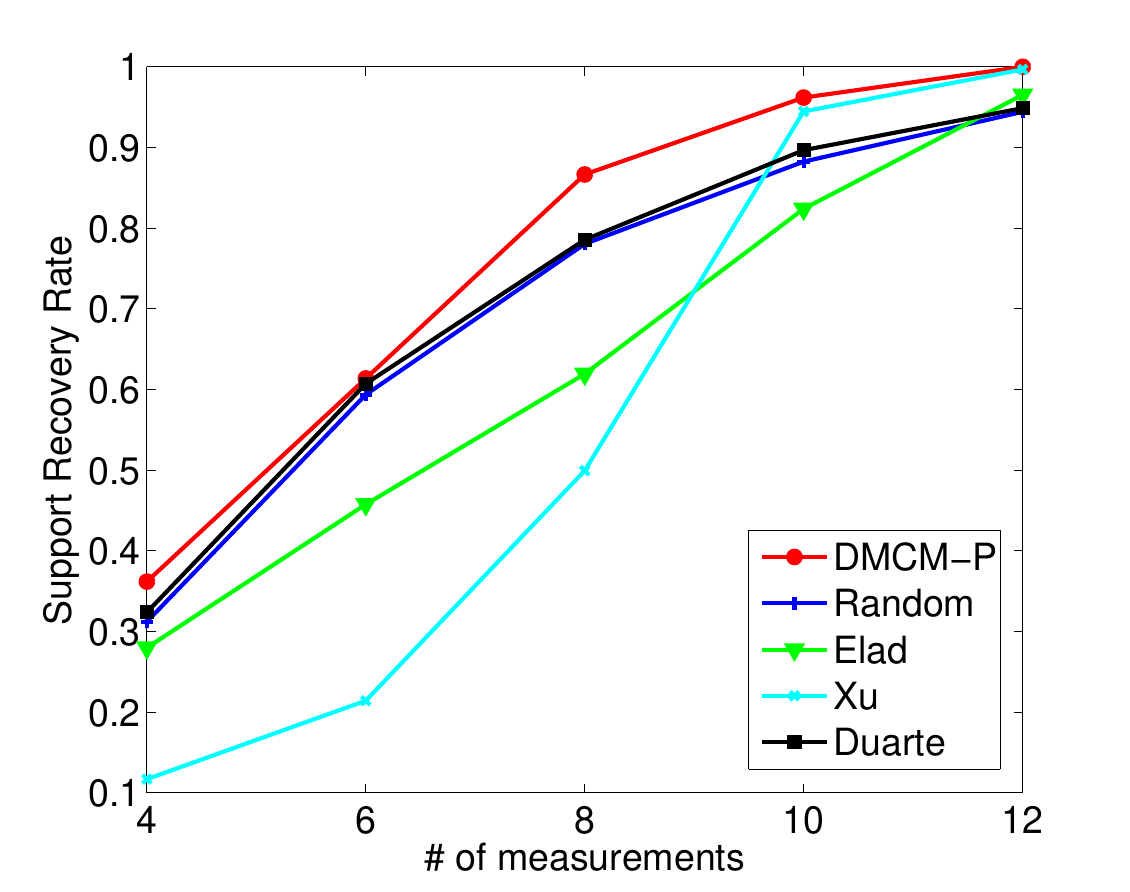}
    \end{tabular}
    \caption{Signal reconstruction error and support recovery rate v.s. number of measurements, where $\D$ is the DCT matrix.}
    \label{fig_cs_dct_measurement}
\end{figure}
\begin{figure}
    \centering
    \begin{tabular}{c@{\extracolsep{0.5em}}c@{\extracolsep{0.5em}}c@{\extracolsep{0.5em}}}
    \includegraphics[width=0.53\textwidth]{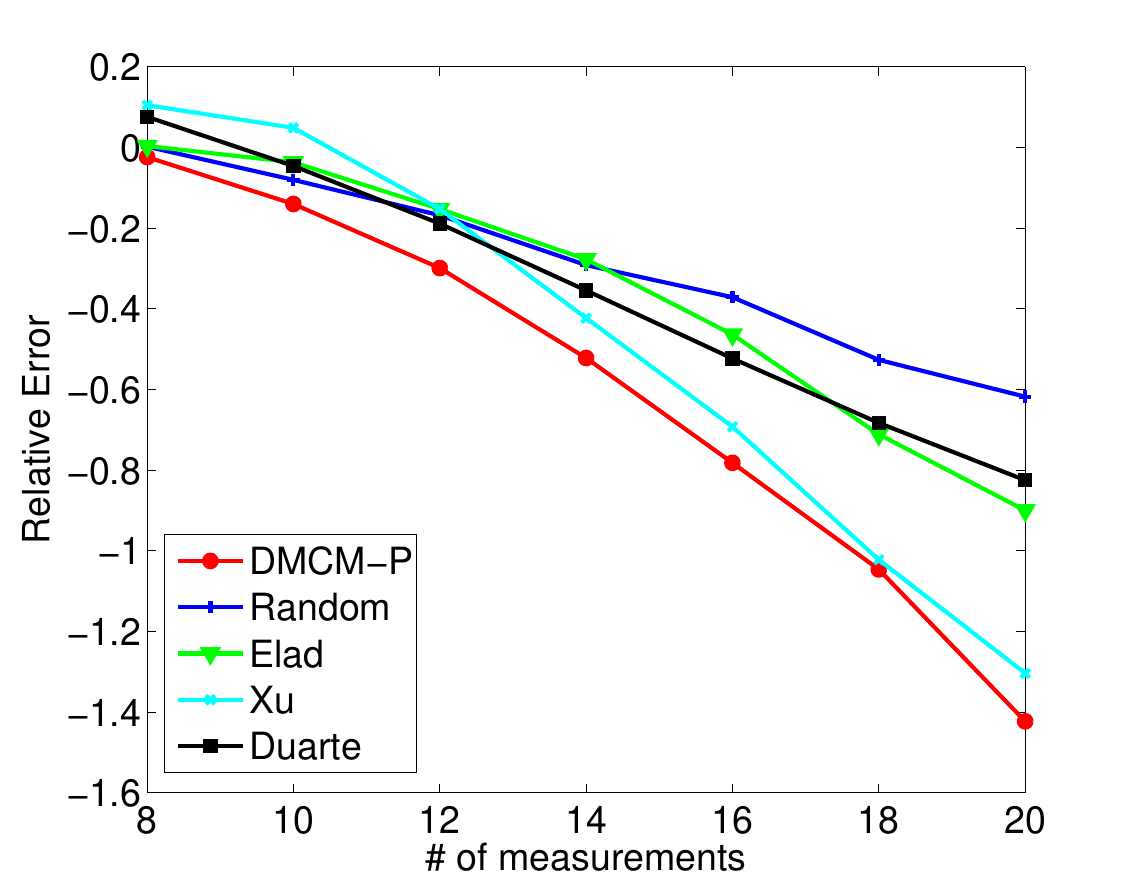}
    &\hspace{-0.6cm}\includegraphics[width=0.53\textwidth]{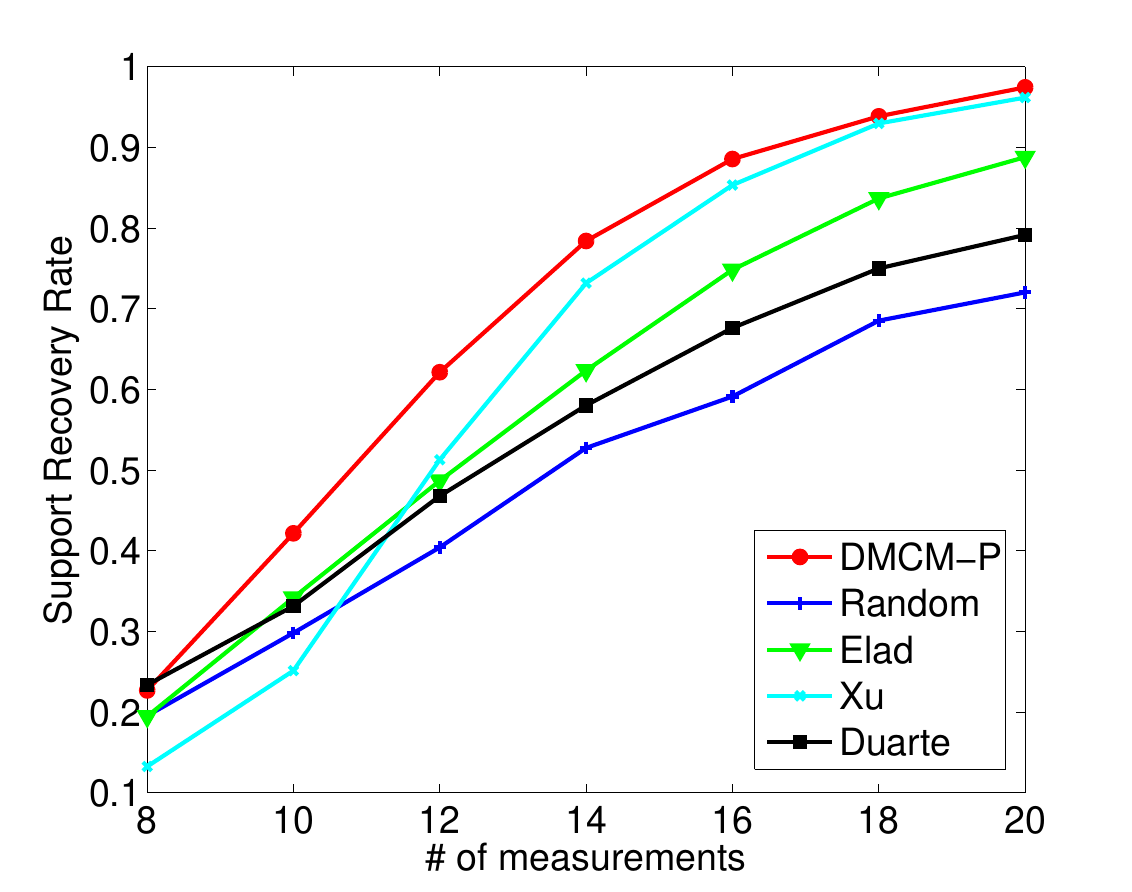}
    \end{tabular}
    \caption{Signal reconstruction error and support recovery rate v.s. number of measurements, where $\D$ is learned by K-SVD.}
    \label{fig_cs_ksvd_measurement}
\end{figure}

In the first experiment, we {\color{black} vary $m$ and set $n=60$, $d=30$, $T=2$ when $\D$ is the Gaussian random matrix, $n=60$, $d=60$, $T=2$ when $\D$ is the DCT matrix and $n=100$, $d=100$, $T=4$ when $\D$ is the matrix learned by K-SVD, respectively. Figure \ref{fig_cs_gaussian_measurement}, \ref{fig_cs_dct_measurement} and \ref{fig_cs_ksvd_measurement} show the average
relative reconstruction error {\color{black}(left) and support recovery rate (right)} v.s. the number $m$ of measurements
($T$ is fixed). In the last case, we follow \cite{Eladbook} to train a dictionary for sparsely representing patches of size 10$\times$10 extracted from the
image Barbara. This image is of size 512$\times$512 and thus has 253009 possible patches, considering all overlaps. We extract one tenth of these patches (uniformly spread) to train on using the K-SVD with 50 iterations.} The CS performance improves as $m$ increases.
Also, as expected, all the optimized projection matrices produce
better CS performance than the random projection does, and our
proposed DMCM-P consistently outperforms the algorithms of Elad,
Xu et al. and Duarte-Carajalino and Sapiro.

\begin{figure}
    \centering
    \begin{tabular}{c@{\extracolsep{0.5em}}c@{\extracolsep{0.5em}}c@{\extracolsep{0.5em}}}
    \includegraphics[width=0.48\textwidth]{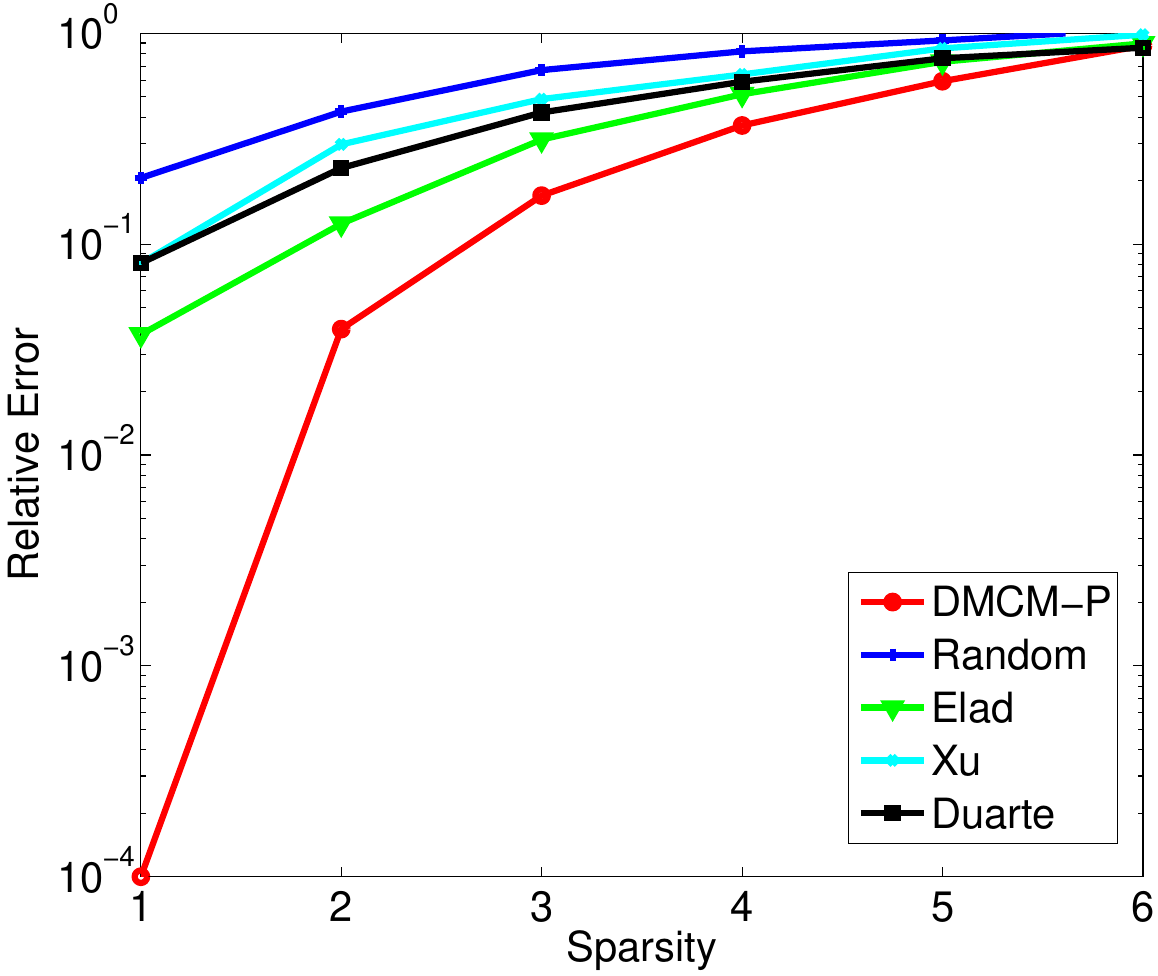}
    &\includegraphics[width=0.53\textwidth]{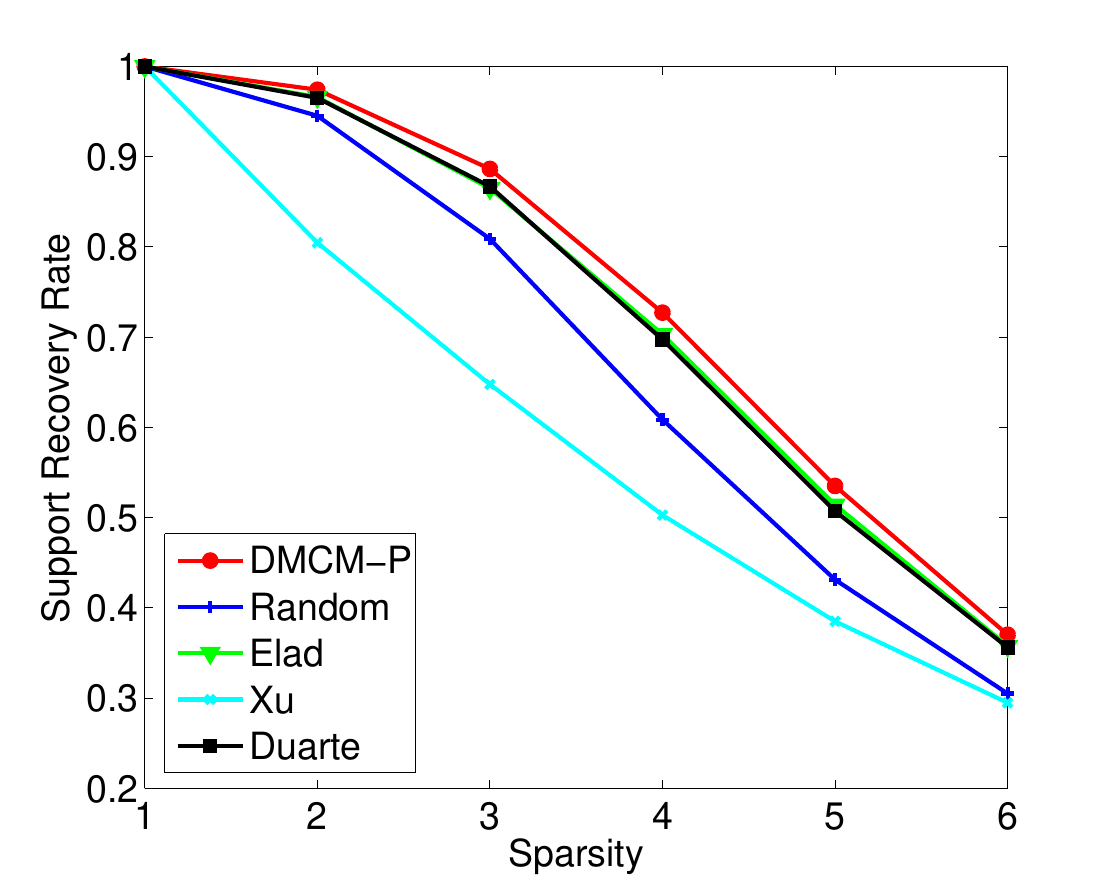}
    \end{tabular}
    \caption{Signal reconstruction error and support recovery rate v.s. sparsity, where $\D$ is the Gaussian random matrix.}
    \label{fig_cs_gaussian_sparsity}
\end{figure}
\begin{figure}
    \centering
    \begin{tabular}{c@{\extracolsep{0.5em}}c@{\extracolsep{0.5em}}c@{\extracolsep{0.5em}}}
    \includegraphics[width=0.53\textwidth]{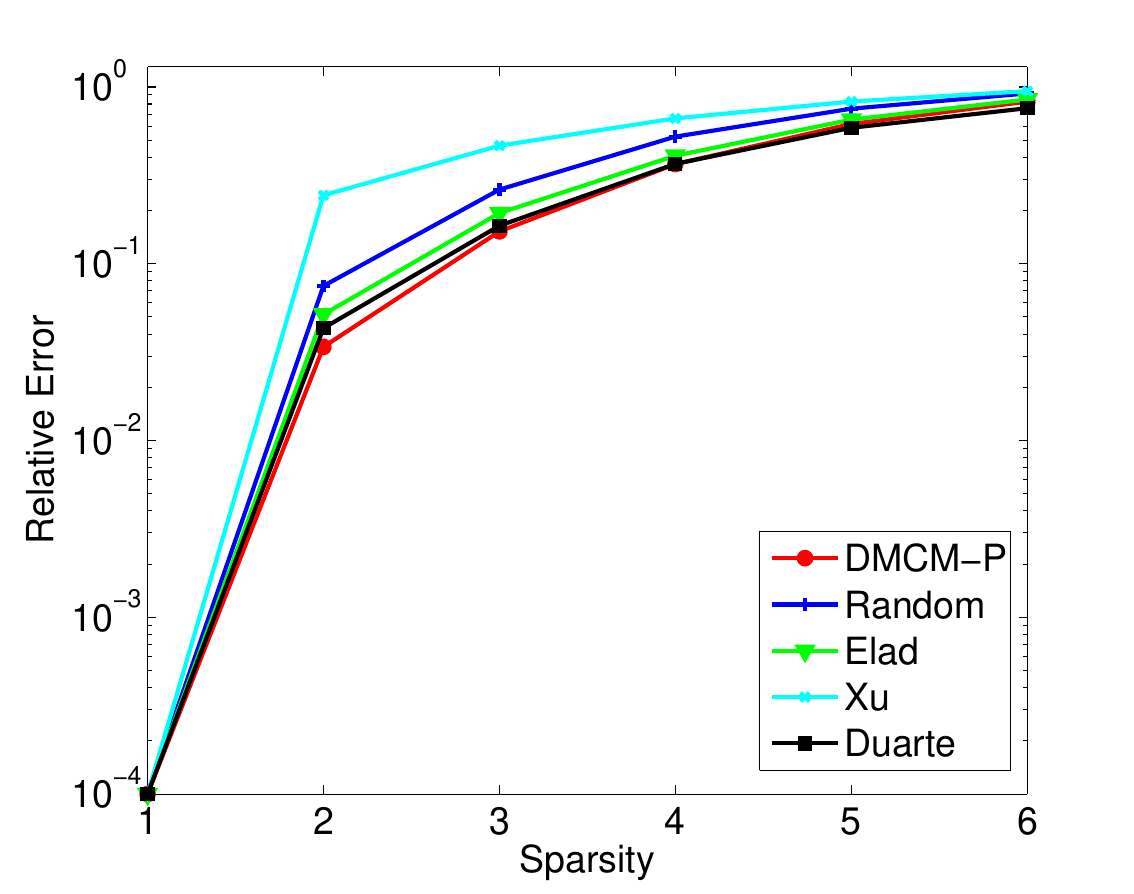}
    &\hspace{-0.6cm}\includegraphics[width=0.53\textwidth]{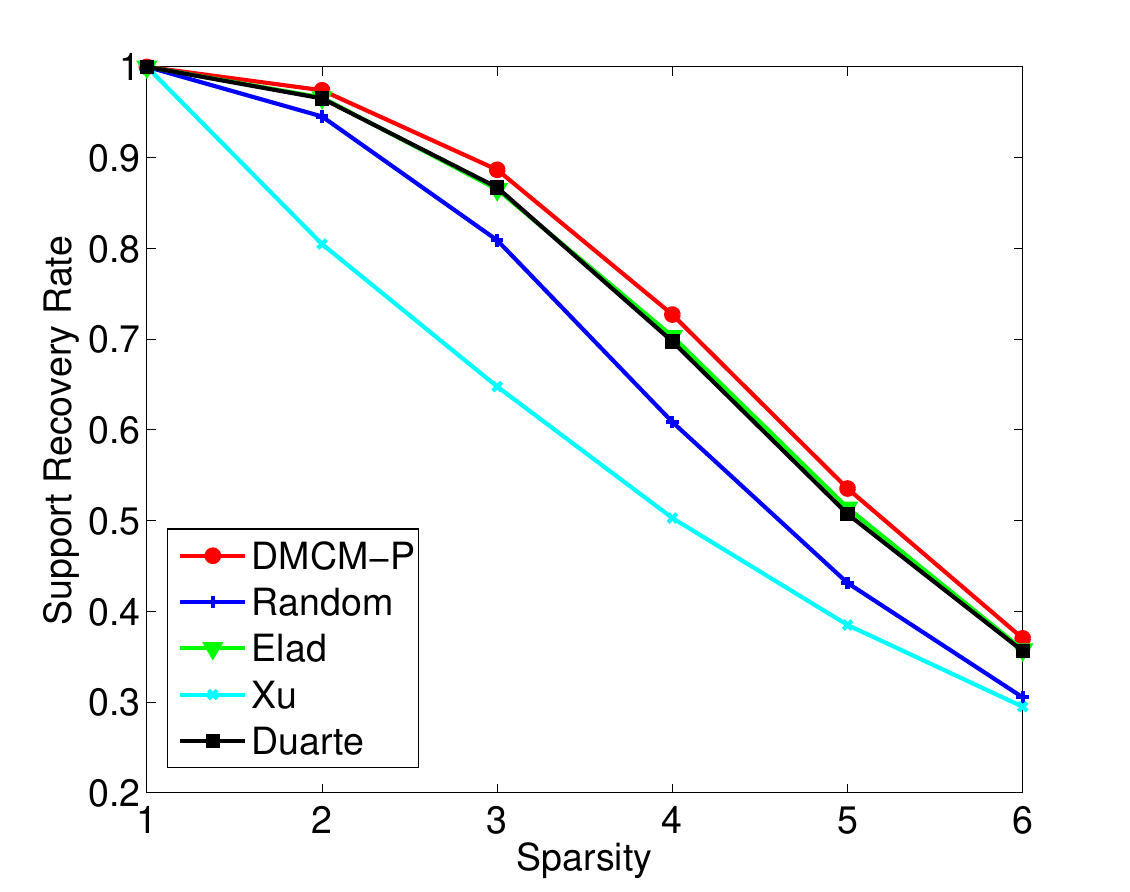}
    \end{tabular}
    \caption{Signal reconstruction error and support recovery rate v.s. sparsity, where $\D$ is the DCT matrix.}
    \label{fig_cs_dct_sparsity}
\end{figure}
\begin{figure}
    \centering
    \begin{tabular}{c@{\extracolsep{0.5em}}c@{\extracolsep{0.5em}}c@{\extracolsep{0.5em}}}
    \includegraphics[width=0.53\textwidth]{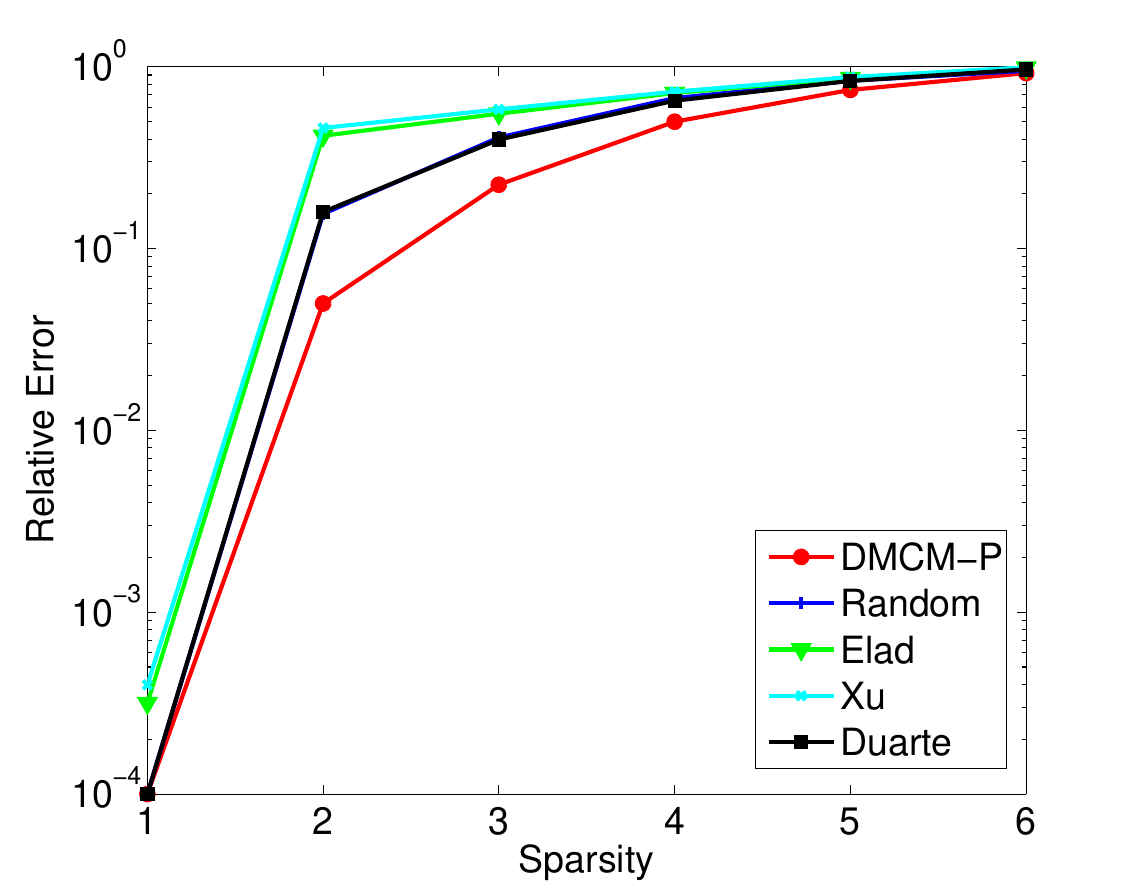}
    &\hspace{-0.6cm}\includegraphics[width=0.53\textwidth]{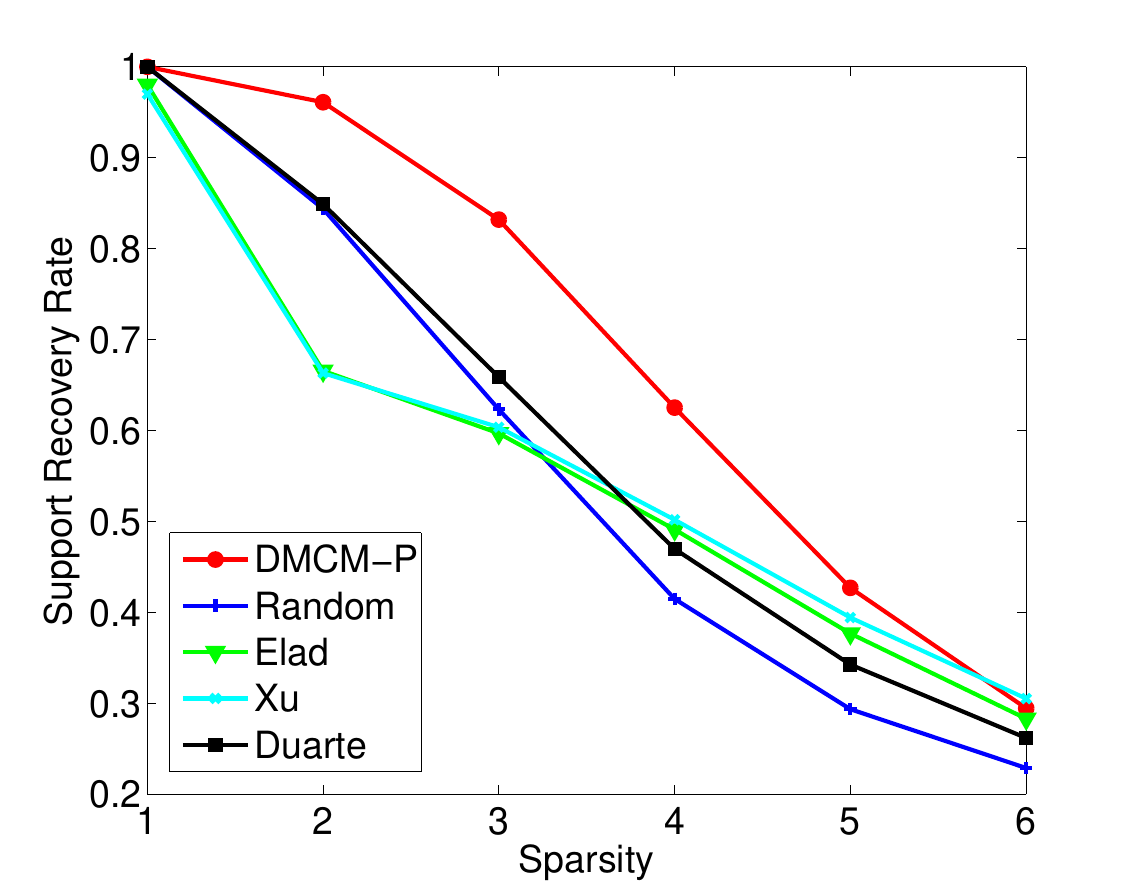}
    \end{tabular}
    \caption{Signal reconstruction error and support recovery rate v.s. sparsity, where $\D$ is learned by K-SVD.}
    \label{fig_cs_ksvd_sparsity}
\end{figure}

In the second experiment, we {\color{black} vary the sparsity level $T$ and set $m=18$, $n=180$ and $d=90$ when $\D$ is the Gaussian random matrix, $m=15$, $n=180$ and $d=180$ when $\D$ is the DCT matrix and $m=12$, $n=100$ and $d=100$ when $\D$ is the matrix learned by K-SVD. Figure
\ref{fig_cs_gaussian_sparsity}, \ref{fig_cs_dct_sparsity} and \ref{fig_cs_ksvd_sparsity} show the average relative
reconstruction error and support recovery rate as a function of the sparsity level $T$ ($m$
is fixed).} The CS performance also improves as $T$ decreases.
Also, our DMCM-P consistently outperforms random projection and
other deterministic projection optimization methods. This is due
to the low mutual coherence of $\PP\D$ thanks to our optimized
projection method as verified in the previous experiments.

\begin{figure}
    \centering
    \begin{tabular}{c@{\extracolsep{0.5em}}c@{\extracolsep{0.5em}}c@{\extracolsep{0.5em}}}
    \includegraphics[width=0.53\textwidth]{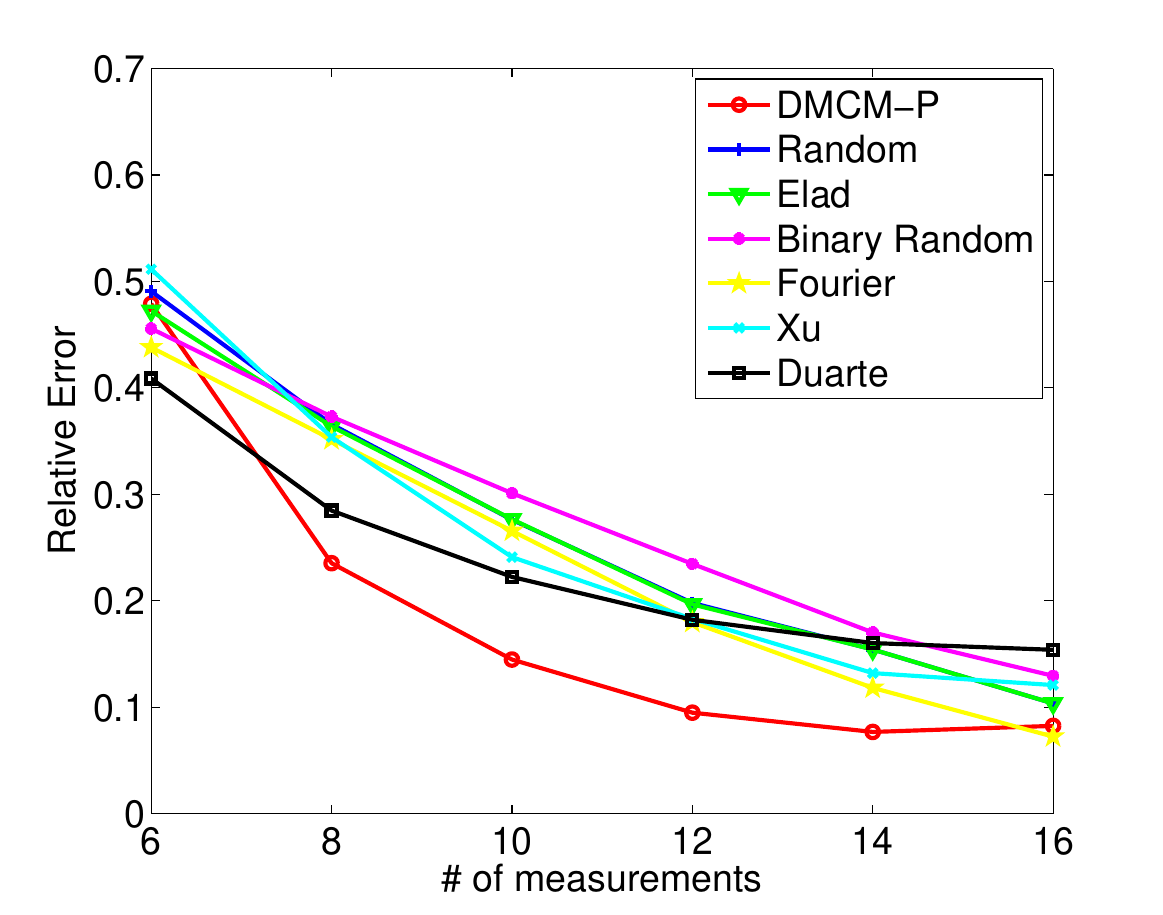}
    &\hspace{-0.6cm}\includegraphics[width=0.53\textwidth]{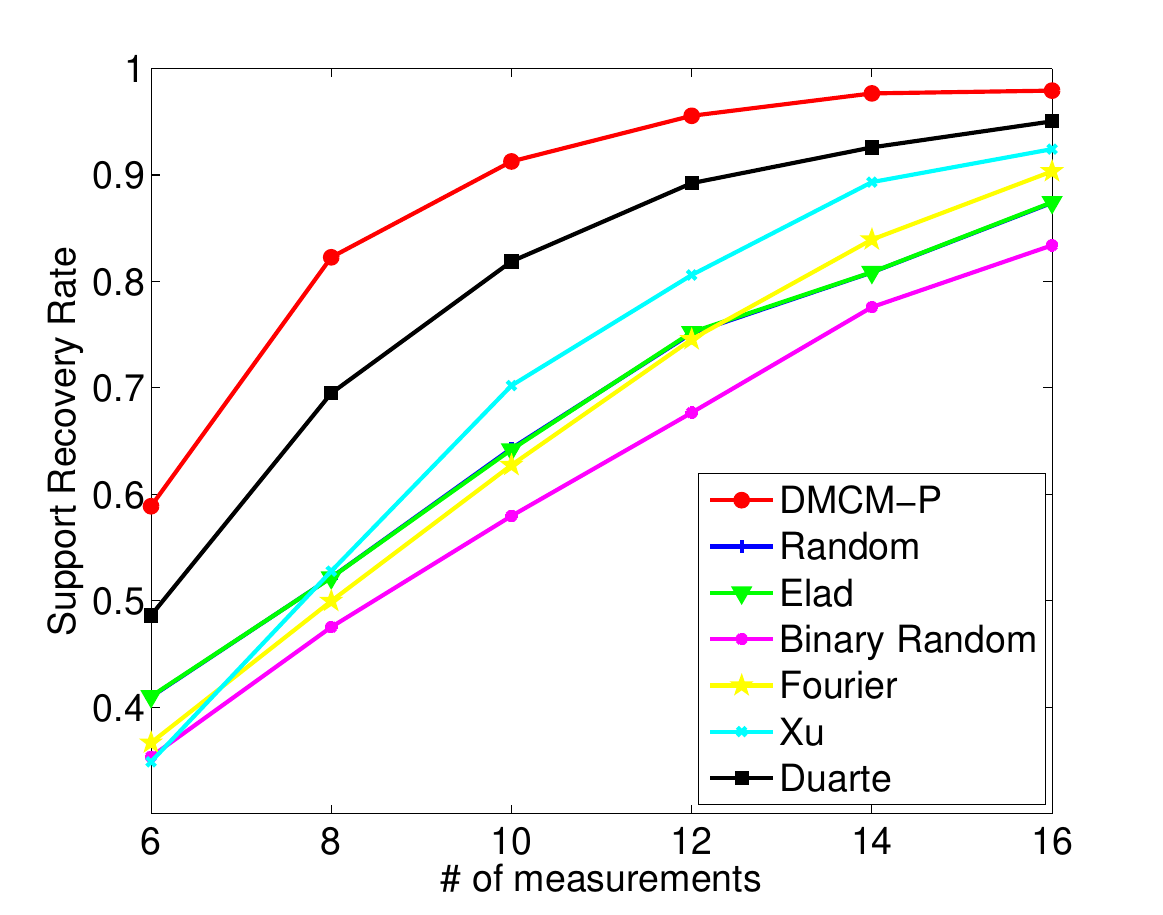}
    \end{tabular}
    \caption{{\color{black} Signal reconstruction error and support recovery rate v.s. measurement in the noisy case, where $\D$ is the Gaussian random matrix.}}
    \label{fig_cs_gaussian_noisy}
\end{figure}

We also test the noisy case. We add Gaussian random noise with 0
mean and 0.01 variance to each element of the observation $\y$ and
then recover the true signal from this noisy $\y$. This time we
test with $\D$ in another different distribution and another
choice of the ratio $n/d$.
%This time we
%test with different distributions of $\D$ and the ratio $n/d$.
We generate elements of $\D$ by a uniform distribution on [0,1]. We
choose $m=[6:2:16]$, $d=40$ and $n=60$. {\color{black} Besides the sensing matrices constructed via optimization, we also compare DMCM-P with the the random binary matrix and Fourier matrix with random selected rows. Figure
\ref{fig_cs_gaussian_noisy} shows the performance comparison based on the relative reconstruction error {and support recovery rate} v.s. the number of measurements. It can be seen that our method also achieves the best performance in almost all cases. The improvement of our method over the random sensing matrices (using Fourier matrix with random selected rows or the random binary matrices) are significant.}

%If the relative error of a reconstruction exceeds $10^{-4}$, the construction is considered to be a failure. To reduce the overall runtime, we follow \cite{elad2007optimized} to stop the test if more than 300 failures were accumulated and the average so far was used instead.

%We also test model (\ref{prolcd}) in compressed sensing. Now we optimize matrix $\M$ directly. $\alpha$ is a $T$-sparse signal with randomly chosen locations and the nonzeros are chosen by a uniform distribution in $[-1,1]$. $\y$ is generated by $\y=\M\alpha$. At last we solve problem (\ref{eq_ed}) by OMP. We test the relative error $\|\alpha-\hat\alpha\|_2$ as a function of the number of measurements $m=[6:2:16]$ with $n=60,d=30,T=2$ and the sparsity level $T=[1:6]$ with fixed $m=16,n=180,d=90$, respectively. Figure \ref{fig_6_cs1measurements} and \ref{fig_7_errorvssparsity} shows the result.

\section{Conclusions}
This paper focuses on optimizing the projection matrix in CS for
reconstructing signals which are sparse in some overcomplete
dictionary. We develop the first model which aims to find a
projection $\PP$ by minimizing the mutual coherence of $\PP\D$
directly. We solve the nonconvex problem by alternating
minimization and prove the convergence. Simulation results show
that our method does achieve much lower mutual coherence of
$\PP\D$, and also leads to better CS performance. Considering that
mutual coherence is important in many applications besides CS, we
expect that the proposed construction will be useful in many other
applications as well, besides CS.

There is some interesting future work. First, though we give the
first solver with convergence guarantee in Algorithm \ref{alg1}
for (\ref{prolcd}), the obtained solution is not guaranteed to be
globally optimal due to the nonconvexity of the problem. It is
interesting to investigate when the obtained solution is globally
optimal. {Second, currently the proposed method is not efficient, and it is valuable to find faster solvers. For example, we may consider solving (\ref{prolcd}) and (\ref{lmcpideal}) by Alternating Direction Method of Multiplier (ADMM) after introducing some auxiliary variables, which may be
more efficient than our current solvers. But proving its convergence for nonconvex problems, (\ref{prolcd}) and (\ref{lmcpideal}), will be challenging.}

\section*{Appendix}
In this section, we give the proof of Theorem
\ref{convergence_p2}.
\begin{definition}\cite{Rockafellar1998,Jerone2013}
Let $g$ be a proper and lower semicontinuous function.
\begin{enumerate}
 \item For a given $\x\in \text{dom} \ g$, the Frech\'{e}t subdifferential of $g$ at $\x$, written as $\hat\partial g(\x)$, is the set of all vectors $\u\in \mathbb R^n$ which satisfies
    \begin{equation}
    \liminf_{\y \neq\x,\y\rightarrow\x} \frac{g(\y)-g(\x)-\left\langle\u,\y-\x\right\rangle}{\|\y-\x\|}\geq 0.\notag
    \end{equation}
    %When $\x\in$dom$g$, we set $\hat\partial g(\x)=\emptyset$.
 \item The limiting-subdifferential, or simply the subdifferential, of $g$ at $\x\in\mathbb R^n$, written as $\partial g(\x)$, is defined through the following closure process
    \begin{eqnarray}
    \partial g(\x)&:=&\{\u\in\mathbb R^n:\exists \x_k\rightarrow \x,g(\x_k)\rightarrow g(\x),\notag\\
    &&\u_k\in\hat\partial g(\x_k)\rightarrow\u,k\rightarrow \infty\}.\notag
    \end{eqnarray}
\end{enumerate}
\end{definition}

\begin{proposition}\label{prop}\cite{Rockafellar1998,Jerone2013}\label{proposition_nonsmooth} The following results hold:
\begin{enumerate}
 \item In the nonsmooth context, the Fermat's rule remains
unchanged: If $\x\in \mathbb R^n$ is a local minimizer of $g$,
then $0\in\partial g(\x)$.
 \item Let $(\x_k,\u_k)$ be a sequence
such that $\x_k\rightarrow\x$, $\u_k\rightarrow\u$,
$g(\x_k)\rightarrow g(\x)$ and $\u_k\in\partial g(\x_k)$. Then
$\u\in\partial g(\x)$.
 \item If $f$ is a continuously
differentiable function, then $\partial(f+g)(\x)=\nabla
f(\x)+\partial g(\x)$.
\end{enumerate}
\end{proposition}

 \vspace{8pt}

\noindent \textbf{Proof of Theorem \ref{convergence_p2}:}
First,  (\ref{updateM2}) can be rewritten as
 \begin{align}
&\M_{k+1} \notag\\
=& \arg\min_{\M} \langle\nabla f_{\rho}(\M_k), \M-\M_k\rangle+\frac{1}{2\alpha}\norm{\M-\M_k}_F^2 \notag\\
&+\frac{1}{2\beta}\|\M-\PP_k\D\|_F^2+h(\M).\notag
\end{align}
By the optimality of $\Mkk$, we have
\begin{align}
&h(\M_{k+1})+\left\langle\nabla f_{\rho}(\M_k),\M_{k+1}-\M_k\right\rangle\notag\\
&+\frac{1}{2\alpha}\|\M_{k+1}-\M_k\|_F^2+\frac{1}{2\beta}\|\M_{k+1}-\PP_{k}\D\|_F^2\notag\\
\leq& h(\M_k)+\frac{1}{2\beta}\|\M_{k}-\PP_k\D\|_F^2. \label{p1_con1}
\end{align}
From the Lipschitz continuity of $\nabla f_{\rho}(\M)$, we have
\begin{align}
&F(\M_{k+1},\PP_k)\notag\\
=&f_{\rho}(\M_{k+1})+\frac{1}{2\beta}\|\M_{k+1}-\PP_k\D\|_F^2\notag\\
\leq& f_{\rho}(\M_{k})+\left\langle\nabla f_{\rho}(\M_k),\M_{k+1}-\M_k\right\rangle\label{p1_con2}\\
&+\frac{1}{2\rho}\|\M_{k+1}-\M_k\|_F^2+\frac{1}{2\beta}\|\M_{k+1}-\PP_k\D\|_F^2.\notag
\end{align}
Add (\ref{p1_con1}) and (\ref{p1_con2}), we have
\begin{align}
&h(\M_{k+1})+F(\M_{k+1},\PP_k)\notag\\
\leq& h(\M_k)+f_{\rho}(\M_{k})-\left(\frac{1}{2\alpha}-\frac{1}{2\rho}\right)\|\M_{k+1}-\M_k\|_F^2\notag\\
&+\frac{1}{2\beta}\|\M_{k}-\PP_k\D\|_F^2\label{eq3111}\\
=&h(\M_{k})+F(\M_{k},\PP_k)-\left(\frac{1}{2\alpha}-\frac{1}{2\rho}\right)\|\M_{k+1}-\M_k\|_F^2.\notag
\end{align}
\canyi{Note that $F(\M_{k+1},\PP)=\frac{1}{2\beta}\|\M_{k+1}-\PP\D\|_F^2$ is $\frac{1}{\beta}\sigma_{\min}^2(\D)$-strongly convex, where $\sigma_{\min}(\D)$ denotes the smallest singular value of $\D$ and it is positive since $\D$ is of full rank. Then by Lemma B.5 in \cite{mairal2013optimization} and the optimality of $\PP_{k+1}$ to (\ref{P_step}), we have
\begin{eqnarray}
F(\M_{k+1},\PP_{k+1})\leq F(\M_{k+1},\PP_{k}) - \frac{1}{2\beta}\sigma_{\min}^2(\D)\norm{\PP_{k+1}-\PP_{k}}_F^2.\label{eq3111222}
\end{eqnarray}
Combining (\ref{eq3111}) and (\ref{eq3111222}) leads to
\begin{align}
&h(\M_{k+1})+F(\M_{k+1},\PP_{k+1})\notag\\
\leq& h(\M_k)+F(\M_k,\PP_k)-\left(\frac{1}{2\alpha}-\frac{1}{2\rho}\right)\|\M_{k+1}-\M_k\|_F^2- \frac{1}{2\beta}\sigma_{\min}^2(\D)\norm{\PP_{k+1}-\PP_{k}}_F^2. \label{boundF}
\end{align}
}

\begin{CanyiPar}{black}
	Second, by the optimality of $\Mkk$, we have
	\begin{align}
	0 \in & \partial h(\M_{k+1})+\nabla f_{\rho}(\M_k)+\frac{1}{\alpha}(\M_{k+1}-\M_k)\notag\\
	 &+\frac{1}{\beta}(\M_{k+1}-\PP_k\D).\label{p1_con3}
	\end{align}
	Thus, there exists $\W_{k+1}\in \nabla_{\M} F(\M_{k+1},\PP_{k+1})+\partial h(\Mkk)$, such that
	\begin{align}
	\W_{k+1}  \in &\nabla  f_\rho(\Mkk)+ \frac{1}{\beta}(\Mkk - \PP_{k+1} \D ) +\partial h(\Mkk) \notag\\
	 = & \nabla  f_\rho(\Mk)+ \frac{1}{\beta}(\Mkk - \PP_{k} \D ) +\partial h(\Mkk) \label{wboundto2}\\
	   & +  (f_\rho(\Mkk)-f_\rho(\Mk)) + \frac{1}{\beta}(\PP_{k} - \PP_{k+1})\D. \notag
	\end{align}
Then, combining (\ref{p1_con3}) and (\ref{wboundto2}) leads to
\begin{align}
 \norm{\W_{k+1}}_F \leq & \norm{\nabla  f_\rho(\Mk)+ \frac{1}{\beta}(\Mkk - \PP_{k} \D )
 	+\partial h(\Mkk)}_F \notag\\
 &+ \norm{f_\rho(\Mkk)-f_\rho(\Mk)}_F + \frac{1}{\beta}\norm{(\PP_{k} - \PP_{k+1})\D}_F \\
 \leq  &\frac{1}{\alpha} \norm{\Mkk - \Mk}_F + \frac{1}{\rho}  \norm{\Mkk - \Mk}_F + \frac{1}{\beta}\norm{\D}^2\norm{\PP_{k} - \PP_{k+1}}_F, \label{boundwondiff}
\end{align}
where (\ref{boundwondiff}) uses the property that $\nabla f_\rho (\M)$ is Lipschitz continuous with the Lipschitz constant  $1/\rho$. Also, by the optimality of $\PP_{k+1}$, we have
\begin{eqnarray}\label{optPPP}
\0=\nabla_{\PP} F(\M_{k+1},\PP_{k+1})=(\M_{k+1}-\PP_{k+1}\D)\D^T.
\end{eqnarray}

Third, note that $F(\M,\PP)$ is coercive, i.e., $F(\M,\PP)$ is
bounded from below and $F(\M,\PP)\rightarrow+\infty$ when
$\|[\M,\PP]\|_F\rightarrow+\infty$. It can be seen from
(\ref{boundF}) that $F(\M_k,\PP_k)$ is bounded. Thus
$\{\M_k,\PP_k\}$ is bounded. Then there exists an accumulation
point $(\M^*,\PP^*)$ and a subsequence $\{\M_{k_j},\PP_{k_j}\}$
such that $(\M_{k_j},\PP_{k_j})\rightarrow (\M^*,\PP^*)$ as
$j\rightarrow+\infty$.
Since $F(\M,\PP)$ is continuously differentiable, we have
$F(\M_{k_j},\PP_{k_j})\rightarrow F(\M^*,\PP^*)$. As $h(\M_k)=0$
for all $k$ and the set $\{\M:\|\M_i\|_2=1,i=1,\cdots,n\}$ is
closed, we have $h(\M^*)=0$ and
$F(\M_{k_j},\PP_{k_j})+h(\M_{k_j})\rightarrow
F(\M^*,\PP^*)+h(\M^*)$.

$\hfill\blacksquare$
\end{CanyiPar}

\noindent{\large{\textbf{Reference}}}

{
	\bibliographystyle{IEEEbib}
	\bibliography{dmcm}
}

\end{document}

%% file: MACPdef-ams.tex
\newcommand{\diagop}{\operatorname{diag}}

\newcommand{\diag}[1]{\ensuremath{\diagop\left(#1\right)}}
\newcommand{\rankop}{\operatorname{rank}}
\newcommand{\rank}[1]{\ensuremath{\rankop\left(#1\right)}}

%% file: MACPdef.tex
\newcommand{\A}{\ensuremath{\mathbf{A}}}

\newcommand{\G}{\ensuremath{\mathbf{G}}}

\newcommand{\I}{\ensuremath{\mathbf{I}}}

\newcommand{\PP}{\ensuremath{\mathbf{P}}}

\renewcommand{\SS}{\ensuremath{\mathbf{S}}}

\newcommand{\W}{\ensuremath{\mathbf{W}}}

\newcommand{\dd}{\ensuremath{\mathbf{d}}}

\newcommand{\x}{\ensuremath{\mathbf{x}}}
\newcommand{\y}{\ensuremath{\mathbf{y}}}

\newcommand{\0}{\ensuremath{\mathbf{0}}}
\newcommand{\1}{\ensuremath{\mathbf{1}}}

\newcommand{\balpha}{\ensuremath{\boldsymbol{\alpha}}}

{%
\begin{list}{#1}{
\vspace{-\topsep}
\vspace{-\partopsep}
\setlength{\itemindent}{0cm}
\setlength{\rightmargin}{0cm}
\setlength{\listparindent}{0cm}
\settowidth{\labelwidth}{#1}
\setlength{\leftmargin}{\labelwidth}
\addtolength{\leftmargin}{\labelsep}
\setlength{\itemsep}{0cm}
}%
}%
{%
\end{list}
\vspace{-\topsep}
\vspace{-\partopsep}
}

{\begin{enumerate}%
}%
{\end{enumerate}}